\documentclass[amsmath,amsfonts,amssymb,groupedaddress]{revtex4-1}
\usepackage[a4paper,top=3cm,bottom=2cm,left=3cm,right=3cm,marginparwidth=1.5cm]{geometry}
\usepackage{amsmath,amsthm,amssymb,amsfonts}
\usepackage{enumitem}
\usepackage{float}
\usepackage{dcolumn}
 \usepackage{physics}
 \usepackage{subcaption}
 \usepackage{graphicx}
 \usepackage{multirow}
\begin{document}
\title{Correlations, Dynamics, and Interferometry of Anyons in the Lowest Landau Level}

\author{Varsha Subramanyan}
\author{Smitha Vishveshwara}

\affiliation{Department of Physics and Institute for Condensed Matter Theory, University of Illinois at Urbana-Champaign, Urbana, IL, 61801-3080, USA}

\begin{abstract}
 In this article, we present a systematic study of quantum statistics and dynamics of a pair of anyons in the lowerst Landau level (LLL), of direct relevance to quasiparticle excitations in the quantum Hall bulk. We develop the formalism for such a two-dimensional setting of two charged particles subject to a transverse field, including fractional angular momentum states and the related algebra stemming from the anyonic boundary condition, coherent state descriptions of localized anyons, and bunching features associated with such anyons. We analyze the dynamic motion of the anyons in a harmonic trap, emphasizing phase factors emerging from exchange statistics. We then describe non-equilibrium dynamics upon the application of a saddle potential, elaborating on its role as a squeezing operator acting on LLL coherent states, and its action as a beam splitter for anyons. Employing these potential landscapes as building blocks, we analyze anyon dynamics in a quantum Hall bulk interferometer. We discuss parallels between the presented LLL setting and other realms, extensively in the context of quantum optics, whose formalism we heavily borrow from, and briefly in that of black hole phenomena.

\end{abstract}
\maketitle
\section{Introduction}


The theory of identical particles has been a cornerstone in our understanding of physical systems since its original formulation by Heisenberg and Dirac. The establishment of the spin-statistics theorem\cite{textbook} and its wide applicability across modern physics stand as a testament to its importance. Bosons and fermions, as viable quantum entities having symmetric and antisymmetric wavefunctions under exchange and their associated physical properties, are at the heart of our understanding of Nature from the sub-atomic to the astronomical scale. This requirement for the wavefunction to be symmetric or antisymmetric is more than a heuristic conjecture  and has been explicitly shown by considering the topology of the configuration space formed by a system of identical particles\cite{Leinaas}. The two-dimensional world, on the other hand, allows for the intriguing possibility of fractional statistics. That is, on exchanging a pair of these particles, anyons\cite{Wilczek}, their common wavefunction acquires an arbitrary complex phase $e^{i\nu\pi}$ \cite{leshouches,Prange}.  If $\nu$ is a real parameter, the particles are Abelian anyons and if characterized by a matrix defined over a topological space, the particles are non-Abelian anyons. The correlations, interference properties, and dynamics of Abelian anyons form the subject of our work.

Anyons have enjoyed age-old appeal, sparked all the more by the discovery of the fractional quantum Hall effect as a plausible platform for their two-dimensional existence \cite{Prange}. Hallmark signatures of fractional charge and possible corroboration of desired interference patterns have been highly suggestive that quantum Hall quasiparticle excitations could indeed possess the desired anyonic traits. In the past decade, the surge of experimental activity in the realms of novel solid state based quantum Hall settings, cold atomic gases, and topological photonic materials have not only provided tremendous prospects for the realization of fractional quasiparticles, they have provided new ways of manipulating and probing quantum Hall state excitations. In light of these developments, our theoretical studies here of lowest Landau level (LLL) quantum Hall anyons are germane to promisingly accessible physical realizations. 

While fractional quantum Hall anyons have been extensively studied as edge state excitations, often within a Luttinger liquid framework, our work targets bulk LLL anyons, thus enabling us to capture full-fledged two-dimensional attributes. We focus on anyons relevant to the best established groundstates and excitations, namely those associated with Laughlin states. Relevant excitations here correspond to pairs of quasiholes excited above the quantum Hall ground state of filling fraction $\nu=1/m$, where $m$ is an odd integer, created, for instance, through the application of an additional flux quantum. The excitation can be shown to carry fractional charge $e/m$ and, through Berry's phase arguments, to possess the fractional statistical parameter $\nu=1/m$ \cite{Haldane,Laughlin,Halperin83,Arovas}. Most prominent to our treatment, a pair of such excitations can effectively be perceived as two particles in vacuum in the lowest Landau level (LLL) endowed with the appropriate charge an statistics\cite{Haldane}. Hence, our setting, while remarkably simple in studying two particles having anyonic boundary conditions under exchange, directly address fundamental issues associated with anyons and striking differences in comparison with their bosonic and fermionic counterparts.

Our pedagogical study here extensively characterizes the setting for a pair of LLL anyons, descriptions of two such localized pairs, their correlation and associated fractional statistical features, and their dynamics in the presence of analytically tractable potential landscapes. Our work builds on Ref.\cite{Sen1,Sen2}, which set the stage for our treatment here for anyon statics, and Ref. \cite{Smitha10} which does so for dynamics by analyzing the behavior of anyon pairs in saddle potentials. Stemming from the non-commutative nature of the LLL projected position space, we emphasize the nature of the Lie algebra corresponding to anyonic boundary conditions. Explicitly employing this $\mathfrak{sp}(2,\mathbb{R})$  algebra enables us to construct coherent-state anyon descriptions and elegant analyses of their static and dynamic properties. 

Coherent states in the LLL are localized in real space, show particle-like semi-classical behaviour, and can be constructed using the symmetries of the system.  We find that they therefore lend themselves easily to the study of anyon dynamics. We study the dynamics brought about by the presence of an oscillator potential and a saddle potential applied over the LLL. In describing these dynamics, we borrow tools and language from quantum optics to show that wavepacket evolution under the influence of the saddle potential is equivalent to the action of a squeeze operator. The coherent state tunnels through the saddle into transmitted and reflected wavepackets. We characterize the split by means of the Husimi Q-function over the LLL and estimate the tunnelling coefficients for a given initial position. These analyses provide a comprehensive study of LLL anyons statics and dynamics and are amenable to wide application. 

Our analyses directly connect with a range of quantum Hall scenarios. In purely bulk settings, localized anyonic quasiparticles can be expected at local potential minima, as indicated through Coulomb blockade measurement of fractional charge in solid state devices \cite{Yacoby,Lahtinen_2011,Venkatachalam}, or through local flux insertion, which in principle, would be possible in cold atomic gases and photonic materials via laser beams \cite{Zhang18415,Paredes,Liu}. Any off-set of such quasiparticles from the potential minimum would result in the dynamics described here in the context of harmonic potentials. Furthermore, dynamics in the saddle potential geometry is implicitly ubiquitous in the quantum Hall bulk. In solid state systems, edge-state tunneling across the Hall bar can be induced via pinched geometries; the tunneling is mediated via saddle potential scattering through the bulk \cite{Halperin87,VanWees,Tekman}. More fundamentally, disorder plays a major role in solid state systems; one of the more relevant quantum effects in the bulk is, in fact, tunneling between equipotential surfaces via saddle potentials, as quantified in models such as the Chalker-Coddington network\cite{Chalker_1988}. In cold atomic and photonic systems, one of the virtues is the high level of tunability and manipulation; current technologies enable the controlled application of potential landscapes on localized excitations and could realize the dynamics and associated fractional statistical signatures predicted here. Moreover, the tractable potential landscapes presented here can form the building blocks and connections between edge and bulk physics for extensively studied quantum Hall geometries for probing quasiparticle properties, such as Mach-Zehnder and Aharonov-Bohm based interferometers  \cite{West, Slingerland,FabPer}. 

Our presentation here has broad scope in connecting with multiple disciplines. As emphasized above, our formalism heavily borrows from parallels with quantum optics. At heart, the non-commutativity of the LLL maps to a one-dimensional quantum problem in phase space. The mapping naturally leads to photons described by conjugate variables such as number and phase, and position and momentum, and harmonic oscillator levels as analogous to LLL degenerate states. In the presence of the saddle potential, the LLL problem maps to that of a quantum particle in one-dimension in the presence of an inverted oscillator potential, a potential associated with decay in a large range of contexts from alpha-particles to quantum chaos to black hole dynamics. In the context of gravitational physics, we offer a glimpse of the connections described in depth in Ref. \cite{shortpaper}. We mention two striking parallels - transmission across the saddle potential as mirroring Hawking-Unruh radiation \cite{Fulling,Davies_1975,Unruh,hawking1975} and hitherto unexplored temporally decaying quasiparticle modes mimicking signature quasinormal modes of black holes \cite{Vishveshwara:1970zz,qnmpmp}. 

In subsequent sections, our comprehensive presentation is as follows.  In Section II, we  introduce the LLL formulation for single- and two-particle situations. In Section III,  we extensively discuss coherent states, their construction, and their unique properties, highlighting the role of fractional statistics. In Section IV, we perform an in-depth analysis of their dynamics in the presence of harmonic and saddle potentials. We show the influence of fractional statistics on the dynamics of the particle, and draw analogies with the functioning of a beam splitter. In Section V, we heuristically discuss interferometry of these bulk coherent states, along interference paths carved out by appropriate geometries. In Section VI, we explicitly present the analogies with quantum optics,  comparing and contrasting the applicability of the tools we borrow. Finally in Sections VII and VIII, we lay out connections between LLL dynamics and various other seemingly disparate areas of physics, including gravitation,  and speculate on immediate and natural future directions.

\section{Landau Level Physics: Single- and Two-particle Formulations}
As the starting point of our studies, we present the formulation for Landau quantization and lowest Landau level (LLL) physics in the instances of single and and two particles. In this section and the next, we closely follow and build on the formalism laid oout in Ref.\cite{Sen1,Sen2}. While our presentation of one-particle physics is standard \cite{Shankar}, it establishes our convention and forms the stepping stones to build up to the two-particle treatment. For two particles, anyonic boundary conditions give rise to key alterations in the underlying LLL structure. 
\subsection{Single Particle Description}
Our setup consists of the standard situation of charged particles in two dimensions subject to a magnetic field $\vec{B}=B\hat{z}$. 
The Hamiltonian for a single particle is thus given by 
\begin{align}
\mathcal{H}=\frac{1}{2m}\Big(\vec{P}-q\frac{\vec{A}}{c}\Big)^2,\label{ham}
\end{align}

where $\vec{P}$ is the momentum of the particle with components $(P_X,P_Y)$, $m$ is the mass, and $q$ the charge. In principle, the charge could be a fraction of that of the electron, as appropriate for effective quasiparticle descriptions. As for its mass, confining ourselves to the lowest Landau level in this work renders it immaterial. With regards to the magnetic field, we pick the symmetric gauge for the associated vector potential, $\vec{A}=\frac{B}{2}(X,-Y)$. In order to diagonalize the Hamiltonian,  it is useful to perform a coordinate transformation involving the momentum and position operators to a different set given by
\begin{align}\begin{split}\hat{Q}=\frac{1}{qB}\Big(cP_X+\frac{qYB}{2}\Big),\textnormal{ }\hat P=\Big(P_Y-\frac{qXB}{2c}\Big)\end{split}\end{align} 
The operators P and Q are also canonically conjugate, respecting the commutator $[\hat P,\hat Q]=i\hbar$.

The Hamiltonian is then given by
\begin{align}
\mathcal{H}=\frac{\hat P^2}{2m}+\frac{1}{2}m\omega_0^2\hat Q^2
\end{align}
 Hence, the Hamiltonian for a two-dimensional charged particle in a magnetic field maps to that of a one dimensional harmonic oscillator of frequency $\omega_0=\frac{qB}{mc}$. The spectrum thus obtained consists of equally spaced energy levels, namely Landau levels. As with the harmonic oscillator, we define a creation-annihilation operator pair, 
\begin{align}
 B=\Big(\frac{m\omega_0}{2\hbar}\Big)^{1/2}\hat Q+i\Big(\frac{1}{2m\omega_0\hbar}\Big)^{1/2}\hat P\textnormal{,  }
 B^\dagger=\Big(\frac{m\omega_0}{2\hbar}\Big)^{1/2}\hat Q-i\Big(\frac{1}{2m\omega_0\hbar}\Big)^{1/2}\hat P
\end{align}
such that $[B,B^\dagger]=1$. The Hamiltonian takes the diagonal form
\begin{align}
\mathcal{H}=\Big( B^\dagger  B+\frac{1}{2}\Big)\hbar\omega_0.
\end{align}

The ground state, or lowest Landau level wavefunction, can be arrived at by solving the equation $ B\ket{0}=0$. The solutions to this equation are of the form $\Psi_0=e^{-ZZ^*/2}u_k$, where the function $u_k$ is analytic in the complex coordinate Z. Therefore, there is an infinite degeneracy in the ground state (and every other Landau level), characterized by a set of orthogonal states of the form $\psi_{0,k}=N_kZ^ke^{-ZZ^*/4}$. Here, the complex coordinate Z is defined as $Z=(X-iY)/\sqrt{2}l_B$, where the magnetic length  $l_B=\sqrt[]{\frac{c\hbar}{qB}}$ sets the characteristic length scale for the system. The label $k$ indicates the angular momentum of each of the degenerate states, which are eigenstates of the angular momentum operator $L=i\frac{\partial}{\partial\theta}$. We also define the coordinates of the guiding center of the particle motion as 
\begin{align}
    R_x=X+\frac{l_B^2}{\hbar}\hat P,\textnormal{ }R_y=Y-\hat Q
\end{align}

While the ladder operators $({B}, B^\dagger)$ are associated with the one-dimensional harmonic oscillator form corresponding to different Landau levels, given the two dimensional nature of the problem, another set of canonically conjugate variables can be defined and are associated with angular momentum. Specifically, we have  variables 
\begin{align}\begin{split}\hat Q'=\frac{1}{qB}\Big(cP_X-\frac{qYB}{2}\Big)=-R_y\textnormal{, }\hat P'=\Big(P_Y+\frac{qXB}{2c}\Big)=\frac{\hbar}{l_B^2}R_x\end{split}\end{align} 

that respect the commutation relation $[\hat P',\hat Q']=i\hbar$, and commute with $\hat P$ and $\hat Q$. Hence, they do not appear in the Hamiltonian. They form an independent different creation-annihilation pair $( A, A^\dagger)$ via the linear combination
\begin{align}
   A=\Big(\frac{m\omega_0}{2\hbar}\Big)^{1/2}\hat Q'-i\Big(\frac{1}{2m\omega_0\hbar}\Big)^{1/2}\hat P'\textnormal{,  }
 A^\dagger=\Big(\frac{m\omega_0}{2\hbar}\Big)^{1/2}\hat Q'+i\Big(\frac{1}{2m\omega_0\hbar}\Big)^{1/2}\hat P' 
\end{align}
such that $[A,A^\dagger]=1$. It can be shown that the angular momentum operator is defined as $\hat L=( A^\dagger A-B^\dagger B)$. Thus any eigenstate of the Hamiltonian is defined by two quantum numbers corresponding to energy level and angular momentum. When projected onto the lowest Landau level, the energy level index $B^\dagger B$ takes the value zero, thus associating the angular momentum of the LLL with $L=A^\dagger A$. Equipped with this standard formulation for the single particle, we now see how it can be generalized to two particles, giving rise to significant modifications for the case of anyons.

\subsection{Two Particle Description}
In the case of two particles of charge $q$ and mass $m$ in a transverse magnetic field, their relative and center-of-mass (COM) coordinates provide a convenient description. The canonical transformation between their original position and momentum coordinates $(\vec{r_1}, \vec{p_1})$ and $(\vec{r_2}, \vec{p_2})$ and those in the COM and relative frames is given by
\begin{align}
\vec{R}&=\frac{1}{2}(\vec{r_1}+\vec{r_2}), \vec{P}=\vec{P_1}+\vec{P_2}\\
\vec{r}&=\vec{r_1}-\vec{r_2}, \vec{p}=\frac{1}{2}(\vec{P_1}-\vec{P_2}).
\end{align}
For anyons that respect Abelian fractional statistics $\psi(\vec{r_1}-\vec{r_2})= e^{i\pi\nu}\psi(\vec{r_2}-\vec{r_1})$, the appropriate boundary conditions become
\begin{align}
\psi(-\vec{R})=\psi(\vec{R}), \psi(-\vec{r})=e^{i\nu\pi}\psi(\vec{r}).\label{eq:bc}
\end{align}
The boundary condition for the relative coordinate will serve as the key to distinguishing two-particle physics from single-particle physics. For fractional statistics considered here, we have $0\leq\nu\leq 1$. For values of $\nu$ being the inverse of an odd integer, the anyon boundary conditions reflect those for Laughlin quasiparticles. The limiting case of $\nu=0$ and $\nu=1$ correspond to bosons and fermions, respectively. 

Turning to the Hamiltonian describing the system, it is given by
\begin{align}
\mathcal{H}=\frac{1}{4m}\Big(P_x+\frac{qYB}{c}\Big)^2+\frac{1}{4m}\Big(P_y-\frac{qXB}{c}\Big)^2+\frac{1}{m}\Big(p_x+\frac{qyB}{4c}\Big)^2+\frac{1}{m}\Big(p_y-\frac{qxB}{4c}\Big)^2
\end{align}
The Hilbert space for the system as a whole therefore decomposes into a product of those of the relative and COM spaces as $H=H_{COM}\otimes H_r$. In both cases, as with the single particle in a magnetic field, the Hamiltonians may be diagonalized by one set of operators analogous to $(B, B^\dagger)$ in the section above, corresponding to inter-Landau level ladder operators. Another set of operators, which we focus on, analogous to $(A, A^\dagger)$ once again correspond to angular-momentum associated operators confined to the lowest Landau level. The treatment for the COM sector completely parallels that of the single particle. 

Anyonic boundary conditions give rise to significant modifications to the underlying angular momentum and algebraic structures in the relative coordinate frame as compared to the single particle. To elaborate, restricting our discussion to lowest Landau level physics, we consider angular momentum states that respect the anyonic boundary condition of Eq.\ref{eq:bc}, denoted as $\ket{k,\nu}$. In the position basis, these wavefunctions have the form \cite{Smitha10}
\begin{align}
\psi_{k,\nu}=N_{k,\nu}z^{2m+\nu}e^{-zz^*/2}. \label{eq:anyon}
\end{align}
Associated with the angular momentum states, we have a set of conjugate operators $(a, a^\dagger)$ analogous to the single particle $(A, A^\dagger)$ operators. Unlike for the single particle case, however, any linear combination of these operators acting on the angular momentum states generate states that do not respect the anyonic boundary conditions. Thus, we are restricted to bilinears of these operators\cite{Hansson,Kjonsberg}. All such operators can be generated by the set
\begin{align}
\hat{J_1}=\frac{1}{4}(a^\dagger a+aa^\dagger)\textnormal{, }
\hat{J_2}=\frac{1}{4}(a^2+{a^\dagger}^2)\textnormal{, }
\hat{J_3}=\frac{i}{4}(a^2-{a^\dagger}^2).\label{gen}
\end{align}
These operators are the generators of the $\mathfrak{sp}(2,\mathbb{R})$ algebra. The Casimir operator for this algebra has the form $\Gamma^2=\hat{J_1}^2-\hat{J_2}^2-\hat{J_3}^2$ and, by definition, commutes with the generators. 
The generators can be combined to form generalized raising and lowering operators, $\hat B_\pm=\hat{J_2}\pm i\hat{J_3}$. These operators act on the angular momentum states as $\hat B_-\ket{k,\nu}=\sqrt[]{k(k+\nu+\frac{3}{4})}\ket{k,\nu}$.

The angular momentum operator is given by $L=\hbar(2\hat J_1 -\frac{1}{2})$. The angular momentum states respect the eigenequation
\begin{align}
L\ket{k,\mu}&=(2k+\nu)\ket{k,\nu}\\
\Gamma\ket{k,\nu}&=\Big(\frac{\nu}{2}+\frac{1}{4}\Big)\Big(\frac{\nu}{2}-\frac{3}{4}\Big)\ket{k,\nu}.
\end{align} 
These states thus carry fractional angular momentum as required to pick up the statistical phase shift upon the encircling of one anyon around the other. 

With regards to physical observables, as with the single particle situation, we may consider variables such as position. To gain information on the two particles, we may once again decompose into the COM and relative basis. For instance, in the LLL, position operators take the form
\begin{align}
X&=\frac{\ell_B}{2}(A+A^\dagger),\textnormal{ } Y=i\frac{\ell_B}{2}(A-A^\dagger)\\
x&=\ell_B(a+a^\dagger),\textnormal{ } y=i\ell_B(a-a^\dagger).
\end{align}
It must be noted, however, that the anyonic boundary condition once again dictates that only bilinears yield well-defined physical quantities. Having formulated the LLL description of the two-particle system, we now turn to describing localized anyons as wavepackets composed of angular momentum states.

To study particle dynamics in this joint space, we employ coherent states. Coherent states in the lowest Landau level effectively parallel those in 1D phase space, and maximally localized in the plane of the guiding center coordinates. They also follow constant energy trajectories of the applied potential, and hence make good models of particle-like behaviour. 

\section{Coherent States}
Coherent states, in standard simple harmonic oscillator settings, are those formed by a superposition of energy eigenstates such that they respect minimum uncertainty \cite{Shankar}. Moreover, the uncertainties are symmetrically distributed between conjugate variables, such as position and momentum. These states are eigenstates of the annhilation operator associated with the ladder of energy eigenstates. These concepts directly carry over to photons, where the ladder corresponds to different photon numbers, and they play a key role in quantum optics. 

In the case of the lowest Landau level, similar coherent states can be defined. However, energy eigenstates are replaced by momentum eigenstates for the symmetric gauge, and the conjugate variables correspond to guiding center coordinates. Parallel to the position and momentum operators in a 1D phase space, the conjugate variables in the LLL have the commutation relation $[X,Y]=[R_x,R_y]=-il_B^2$. Therefore, the square of the length scale plays the role an effective Planck's constant in the phase space analogy.

Thus, these coherent states form in-plane localized wave-packets respecting minimum uncertainty and their dynamics most closely mimic semi-classical particle behaviour. Here we present the rudimentaries of single-particle coherent states, extensions to anyonic coherent states, and analyses of anyonic coherent state statistical behavior. 
\subsection{Single Particle Coherent States}
In exploring the dynamics of the coherent states, we restrict the system to the LLL. This simply implies a large magnetic field that ensures that the system remains in the ground state configuration. In so restricting the system, we can define coherent states in the LLL as eigenfunctions of the  $A$ operator associated with the LLL angular momentum states, as defined in the previous section. That is, we demand that the action of the operator on the coherent state satisfies the following.
\begin{align}
A\ket{z}_c=z\ket{z}_c
\end{align}
As commonly known\cite{Shankar}, such coherent states have the form
\begin{align}
\ket{z}_c=e^{|z|^2/2}\sum\limits_{k=0}^\infty \frac{z^k}{\sqrt[]{k!}}\ket{k}.\label{eq:coherent}
\end{align}

The average position of any given state is located at complex coordinate $z$. 
These states respect the minimum uncertainty prescribed by the LLL projection in a spatially symmetric manner. Namely, we can consider the uncertainty along the canonically conjugate variables $(X,Y)$, given by $\Delta X = \sqrt{\langle X-\langle X\rangle \rangle^2} $ and $\Delta Y = \sqrt{\langle Y-\langle Y\rangle \rangle^2} $ . Expectation values of these uncertainties in these coherent states respect
\begin{align}
    \langle\Delta X\rangle_z=\frac{l_B}{\sqrt{2}},\textnormal{ } \langle\Delta Y\rangle_z=\frac{l_B}{\sqrt{2}}\implies     \langle\Delta X\rangle_z    \langle\Delta Y\rangle_z=\frac{l_B^2}{2}\label{eq:unc}
\end{align}

Thus, coherent states saturate the minimum bound for LLL projected states given by Eq.\ref{eq:unc}. 

For completeness, to more explicitly and generally discuss LLL projection and the behavior of coherent states, the projection is valid for large enough magnetic field such that the Landau level spacing is much greater than any applied potential $V(x,y)$. Equivalently, this corresponds to the limit $m\rightarrow 0$. It can be shown that the action in the lowest Landau level then takes the form
\begin{align}
\mathcal{S}&=\int dt\Big[\frac{qB}{c}X\dot{Y}-V(X,Y)\Big]\end{align}
We see that the applied potential essentially behaves like the Hamiltonian of the system where the canonically conjugate variables are $\frac{qB}{c}X\equiv\frac{\hbar}{l_B^2}R_x$ and $Y\equiv R_y$. That is, the kinetic degrees of freedom for states in the LLL are frozen and the lowest Landau level acts as a non-commutative plane. Dynamics defined here is therefore analogous to the dynamics of a one dimensional particle in phase space. Further, we can define a group velocity by writing down the semi-classical equations of motion from the action, yielding
\begin{align}
\frac{qB}{c}\dot{Y}=\frac{\partial V}{\partial X}, \textnormal{  }\frac{qB}{c}\dot{X}=-\frac{\partial V}{\partial Y}
\end{align}

Turning to coherent states, their minimum uncertainty in position renders them to most closely resemble particle-like behavior. Moreover, it can be shown that the average position of a coherent state traces equipotential contours along a given applied potential. Such dynamics will prove important in subsequent discussions. 

\subsection{Two-particle Coherent States}

While the treatment of single-particle coherent states is standard, the two-particle coherent state formulation is much more subtle, particularly for the case of anyons\cite{Kjonsberg,Hansson}. To summarize crucial properties, the two-particle coherent state associated with positions $z_1$ and $z_2$ has the product form composed of single-particle states denoted by $\ket{z_1,z_2}_c=\ket{z_1}_c\otimes\ket{z_2}_c$. 
 If the particles are distinguishable, this description is complete. Each individual coherent state behaves like a single-particle state. If the particles are indistinguishable, then the coherent state wavefunction should embody the exchange conditions we demand of them. That is, the coherent states must respect the appropriate boundary conditions characterized by Eq.\ref{eq:bc} and the associated Lie group symmetries. 

As one might expect, the separation of the Hamiltonian into COM $(Z)$ and relative coordinates $(z)$ in the previous section implies that the coherent states themselves are represented in these coordinates. Thus, an equally valid representation of the two particles has the separable form
\begin{align}
\ket{Z,z}_c=\ket{Z}_c\otimes\ket{z}_c
\end{align}
The center of mass coherent state can simply be represented by a regular one-particle state. The relative coordinate coherent state, however, encodes particle statics via the statistical boundary condition. To gain some intuition, let us consider the simplest cases of bosonic and fermionic relative states, referred to as "cat" states in quantum optics literature \cite{DODONOV1974597}. If $\ket{\alpha}_c$ indicates the distinguishable particle coherent state, then these states can be denoted as
\begin{align}
\ket{\alpha_b}_c= N_b (\ket{\alpha}_c+\ket{-\alpha}_c)\\
\ket{\alpha_f}_c=N_f (\ket{\alpha}_c-\ket{-\alpha}_c)
\end{align}
where $N_b,N_f$ are normalization factors, which can be easily evaluated. Compared to the coherent state experssion in Eq.\ref{eq:coherent}, we can thus see that bosonic states are superpositions of even angular momentum states, while fermionic states are superpositions of odd angular momentum states. A logical postulate for the relative coherent state form for particles having fractional statistics is thus

\begin{align}
\ket{z_\nu}_c=N\sum_k \frac{(z/2)^{2k+\nu}}{\sqrt[]{\Gamma(2k+\nu+1)}}\ket{k,\nu}. \label{asymp}
\end{align}
Here, $\ket{k,\nu}$ is the generalized angular momentum state discussed in the previous section for particles having fractional statistics characterized by the parameter $\nu$, where $\nu$ ranges from $0$ for bosons to $1$ for fermions. This state indeed mimics the behavior of a coherent state at large distances compared to the magnetic length. To obtain an exact form of the coherent state, we recall the anyon boundary conditions and the fact that they are respected by quadratic operators formed by the relative coordinate. 

We demand that the coherent state be an eigenvector of the lowering operator of the associated Lie algebra defined before, $B_-=\frac{a^2}{2}$
\begin{align}
B_-\ket{\beta,\nu}_c=\beta\ket{\beta,\nu}
\end{align}
Putting these two together gives us a definition for a generalized coherent state in this algebra -
\begin{align}
\ket{\beta,\nu}_c=N_\beta\sum\limits_k\frac{\beta^k}{\sqrt[]{k!\Gamma(k+\nu+\frac{1}{2})}}\ket{k,\nu}\label{eq:anyonco}
\end{align}
where the normalization factor $N_\beta$ is evaluated to be $N_\beta=\sqrt{\frac{|\beta|^{\nu-\frac{1}{1}}}{I_{\nu-\frac{1}{2}}(2|\beta|)}}$.

Upon comparing with the asymptotically accurate state, it is clear that $\beta$ corresponds to $\frac{1}{2}z^2$.
Equipped with the formalism for coherent state anyons, we now explore their physical properties. 
. 
 
\subsection{Bunching Parameters}
Statistical correlations between particles are manifest in a variety of ways from (anti-)bunching properties in beam splitters and interferometrs to high-energy cross sections to Hanbury-Brown Twiss correlations from the microscopic to astronomical realm \cite{Baym:1997ce, Henny296,Jeltes2007}. Here we focus on a simple measure, which is at the heart of several of these phenomena, the bunching parameter\cite{Baym:1997ce,griffiths_schroeter_2018}. This parameter measures the difference between the average the squared of the separation of two identical quantum particles in a given state and that of two distinguishable particles. Typically, the quantity is positive for fermions, which tend to anti-bunch, and negative for bosons, which tend to bunch. 

It is telling to employ this bunching parameter for studying coherent state anyons, as also studied in the LLL coherent state anyon context in Ref. \cite{Smitha10}. 
We define the parameter as follows\cite{Smitha10}-
\begin{align}
\chi=\frac{1}{4\ell^2}\Big[{_c\bra{\beta,\nu}} \hat{r}^2\ket{\beta,\nu}_c-{_c\bra{z_d}} \hat{r}^2\ket{z_d}_c\Big].\label{bp1}
\end{align}
To evaluate the expectation values, we express the the separation in terms of the angular momentum operators. As shown in the previous section, their action on the angular momentum states is known, thus enabling us to evaluate the bunching parameter for coherent states. Specifically, we have $\hat{r}^2=8\ell^2\hat{J_1}$. Using this expression, we find that $_c\bra{z_d} \hat{r}^2\ket{z_d}_c=(|z|^2+2)\ell^2$. For the anyonic contribution, we have
\begin{align*}
8\ell^2{_c\bra{\beta,\nu}} \hat{J_1}\ket{\beta,\nu}_c&=8\ell^2N_\beta{_c\bra{\beta,\nu}}\sum\limits_{k=0}^\infty\frac{\beta^k}{\sqrt[]{k!\Gamma(k+\nu+\frac{1}{2})}}\hat{J_1}\ket{k,\nu}\\
&=8\ell^2N_\beta^2\sum\limits_{k=0}^\infty\frac{k|\beta|^{2k}}{k!\Gamma(k+\nu+\frac{1}{2})}+4\ell^2(\nu+\frac{1}{2})
\end{align*}
Upon relating the summation above to modified Bessel function identities, we obtain for the bunching parameter in Eq. \ref{bp1}
\begin{align}
\chi&=2|\beta|\frac{I_{\nu+\frac{1}{2}}(2|\beta|)}{I_{\nu-\frac{1}{2}}(2|\beta|)}-2|\beta|+\nu\\
&=A_\phi-2|\beta|+\nu\label{bpe}
\end{align}
Here, the term $A_\phi$ is the Berry's connection associated with the anyonic statistical parameter attributed to the coherent state \cite{Hansson}.
\begin{figure}
\centering
\includegraphics[scale=0.75]{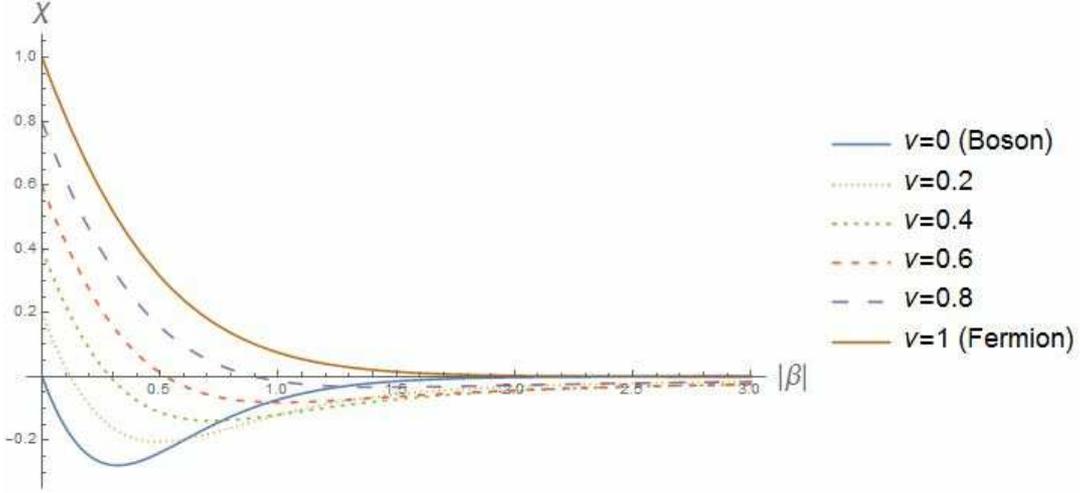}
\caption{Plot of the bunching parameter calculated in Eq.~\ref{bpe} as a function of the square of the separation $|\beta|\sim\frac{|z|^2}{2}$. The parameter measures the average separation of two coherent state quantum particles in comparison with a pair of distiguishable coherent state. Different curves correspond to various values of the statistical parameter $\nu$, with $\nu=0$ corresponding to bosons and $\nu=1$  to fermions. The bunching parameter is clearly always positive for fermions, always negative for bosons, and lies in the intermediate range for anyons, with a sign flip occurring as particle separation increases. For large separations, the parameter vanishes, indicating that the particles are independent and behave like distinguishable particles. }\label{fig:bp}
\end{figure}
The Fig.\ref{fig:bp} is a plot of the bunching parameter as a function of the parameter $|\beta|\sim\frac{|z|^2}{2}$, which represents average separation between the two particles. We see that the bunching parameter is always positive for identical fermions and always negative for identical bosons. Such fermions are more separated on average compared to distinguishable particles and bosons less so. The behaviour of identical anyons lies between these extremes, with the particles showing anti-bunching behaviour for small separations and bunching behaviour for larger separations. The anyons thus morph in nature from fermion-like to boson-like. At large enough distances,  statistical correlations die out and the bunching parameter is uniformly zero for all particles. This plot is comparable to the one obtained in Ref.\cite{Smitha10} where the bunching parameter is calculated for the states in Eq.\ref{asymp}. These states are asymptotically equivalent to $\ket{\beta,\nu}$.
Having analyzed a key static property of anyons, we now study the dynamic manner in which they behave in the presence of applied potentials. 

\section{Coherent state dynamics in the presence of external potentials}

In the absence of a potential landscape, states formed from projecting to the lowest Landau level are frozen due to all states in this space being degenerate. A shallow potential whose strength is much smaller than the Landau level spacing induces dynamics while keeping the projection intact. Here, we demonstrate the manner in which particles endowed with different statistics dynamically respond to such potentials differently. We focus on single-particle and two-particle coherent states described above. We start with the simple example of a harmonic trapping potential, which serves to elucidate distinct features of quantum statistics and then analyze a saddle potential, which can act as a beam splitter. 

\subsection{Harmonic Potential}
Consider an azimuthally symmetric harmonic potential of the form $H_\omega = \lambda(X^2+Y^2)$, first for the case of the single particle. Compared to the single particle purely in the presence of the magnetic field, described by Eq.\ref{ham}, the Hamiltonian takes the form
\begin{align}
\begin{split}
H&=\mathcal{H}=\frac{1}{2m}\Big(\vec{p}-q\frac{\vec{A}}{c}\Big)^2+\lambda(X^2+Y^2)
\end{split}
\end{align}

In the lowest Landau level, the projected Hamiltonian (ignoring the ground state energy) thus takes the form $H_{\lambda,LLL}= \lambda l_B^2(2A^\dagger A+1)$. 
Now considering a single-particle coherent state centered at an initial position $z(0)$, the projected Hamiltonian results in the following time evolution: 

\begin{align}
\begin{split}
\ket{z(t)}_c&=e^{-iH_{\lambda,LLL}t/\hbar}\ket{z(0)}_c\\
&=\sum_{k=0}^{\infty}\frac{z(0)^k}{\sqrt[]{k!}}e^{-it\lambda l_B^2(2A^\dagger A+1)}\ket{k}\\
&=e^{-i\lambda l_B^2t}\sum_{k=0}^{\infty}\frac{(z(0)e^{-it\lambda l_B^2})^k}{\sqrt[]{k!}}\ket{k}\\
&=e^{-i\lambda l_B^2t}\ket{z(0)e^{-it\lambda l_B^2}}_c.
\end{split}
\end{align}

This shows that the coherent state remains coherent, moves in a circle and picks up an additional overall phase. It also reaffirms our expectation that the center of the LLL coherent state follow equipotential lines.

For two particles, given that the harmonic potential is quadratic, we can once again describe the system in center of mass (COM) and relative coordinates. The Hamiltonian in this basis takes the form
\begin{align}
H_{\omega}=2\lambda(X^2+Y^2) + \frac{\lambda}{2}(x^2+y^2). 
\end{align}

As described in Sec.III, LLL coherent states centered at two locations can also be expressed as single particle coherent states $\ket{Z_{com}}_c$ in the center of mass, and $\ket{\beta,\mu}_c$ in the relative coordinate bases.
Due to the separability of the harmonic potential, we can time-evolve the coherent state in separatrely in COM and relative coordinates, since they are independent of each other. For the COM coordinate, this is identical to the single particle case. In the relative coordinate, we have 
\begin{align}
\begin{split}
\ket{\beta(t),\nu}_c&=e^{-iHt/\hbar}\ket{\beta(0),\nu}_c\\
&=\sum_{k=0}^{\infty}\frac{\beta(0)^k}{\sqrt[]{k!\Gamma(k+\nu+\frac{1}{2})}}e^{-i\lambda l_B^2\hat{J_1}}\ket{k,\nu}\\
&=\sum_{k=0}^{\infty}\frac{\beta(0)^k}{\sqrt[]{k!\Gamma(k+\nu+\frac{1}{2})}}e^{-it\lambda l_B^2(k+\frac{\nu}{2}+\frac{1}{4})}\ket{k,\mu}\\
&=e^{-it\lambda l_B^2\frac{\nu}{2}}\sum_{k=0}^{\infty}\frac{\beta(0)^k}{\sqrt[]{k!\Gamma(k+\nu+\frac{1}{2})}}e^{-it\lambda l_B^2(k+\frac{1}{4})}\ket{k,\mu}
\end{split}
\end{align}
 Here, it is important to stress that the parameter $\beta$ is proportional to the square of the position coordinate. Describing a particle exchange in the relative coordinate $z$ space corresponds to a full circle in $\beta$ space. 
 
 The dynamics described here can serve to demonstrate physical processes corresponding to exchange. For instance, in the case of one particle at the bottom of the well at $r_1=0$ and another at some initial radius and azimuthal angle $(r_2, \phi=0)$, the initial COM and relative coordinates are at positions $(r_2/2, 0)$ and $(r_2, \pi)$, respectively. Exchange corresponds to time-evolution such that the second particle is located at $(r_2, \pi)$, corresponding to $\lambda l_B^2 t=2\pi$. In this case, as desired, the COM coordinate does not gain a phase while the relative coordinate coherent state gains a phase of $\sim\pi\nu$. We have thus demonstrated an explicit case of LLL anyons picking up a fractional statistical phase factor upon encircling, as for instance, phenomenologically incorporated into proposals for Aharonov-Bohm based fractional quasiparticle interferometry in quantum Hall systems \cite{West, Slingerland,FabPer}. 

\subsection{Saddle Potential}
\begin{figure}
    \centering
    \includegraphics[width=0.45\textwidth]{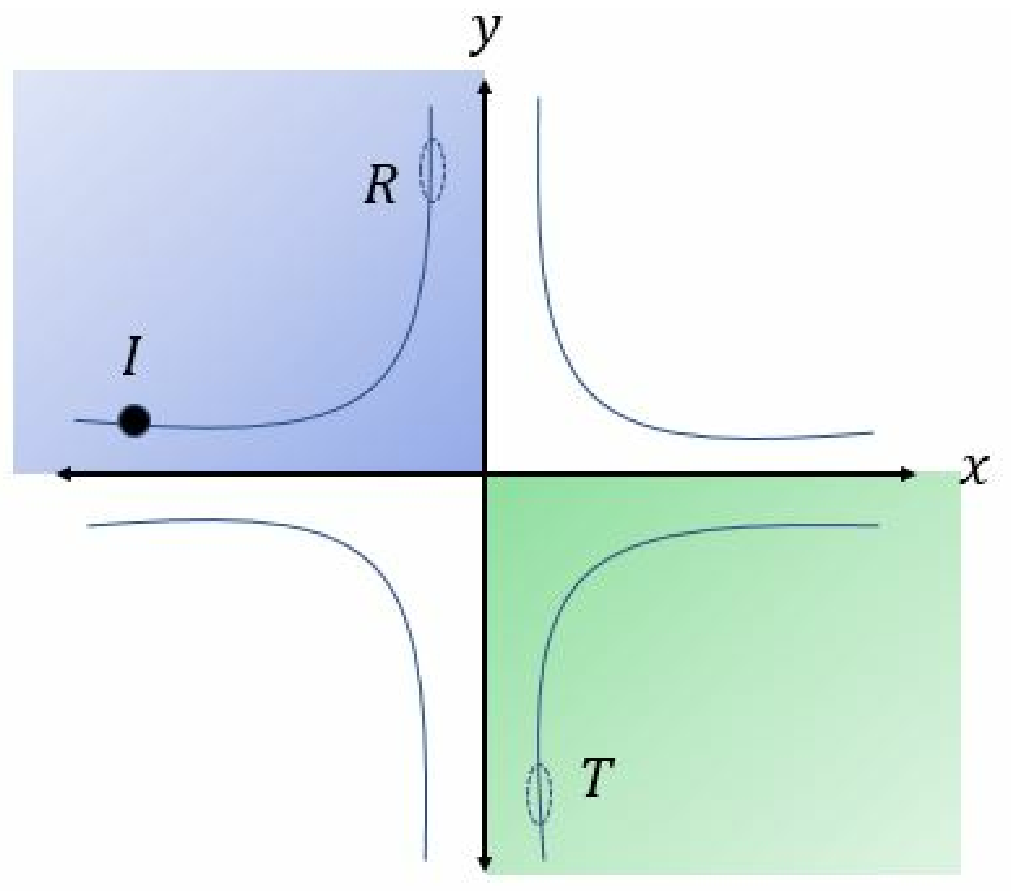}
    \includegraphics[width=0.45\textwidth]{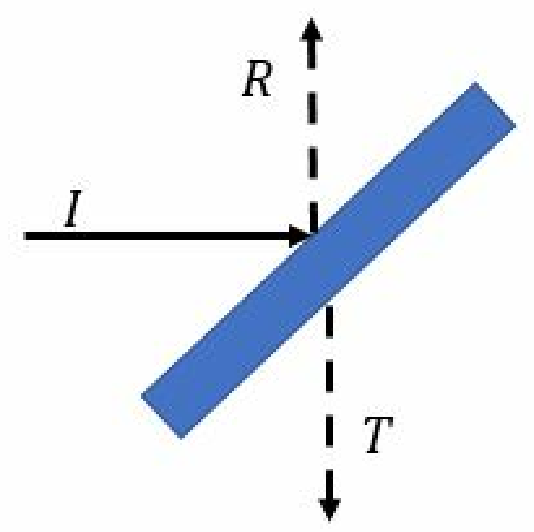}
    \caption{(a) Depiction of the projected saddle potential in the LLL and the trajectory of a localized coherent state in this potential landscape. Upon time evolution, the state becomes squeezed as it travels along an equipotential line and tunnels through the saddle point to give a reflected wavepacket (on the same equipotential trajectory as the initial state) and a transmitted wavepacket (on the opposite equipotential trajectory as the initial state). (b) This action of the saddle potential effectively renders it a beam splitter.}
    \label{fig:saddlepot}
\end{figure}
We now turn to the other instance of a quadratic potential where we can perform an analysis of coherent state evolution and anyonic statistical dependence - the saddle potential. We borrow and build on concepts presented in Ref. \cite{Smitha10, Stone}.  The saddle potential plays a prominent role in quantum Hall physics. It is the relevant feature for tunneling between equipotential surfaces in disordered LLL landscapes. In the case of tunneling between edges across a Hall bar, as commonly brought about by pinched point contact geometries, quasiparticles tunnel through the bulk via saddle potentials\cite{Buttiker90, VanWees}.  

 At heart, the saddle potential can be envisioned as a beam splitter, as for instance, commonly employed in quantum optics, quantum Hall settings, and electron optics. As an illustrative example,  in Fig.\ref{fig:saddlepot}, in a particular configuration of the saddle potential, the incoming wavefunction is primarily along the $x$-axis and transmitted and reflected portions of the wave function travel along opposite directions of the y-axis.
As with any beam splitter, here too, incoming and outgoing modes are related via a scattering matrix (S-matrix) of the form
\begin{align}
\begin{pmatrix}
t&ir\\ir&t
\end{pmatrix}
\end{align}
where $T=|t|^2$ and $R=|r|^2$ are transmission and reflection coefficients, respectively.Unitarity demands the relationships. 
\begin{align}
|t|^2+|r|^2=1, \textnormal{ }
rt^*+tr^*=0
\end{align}
For a 50-50 beam splitter, for example, the S matrix assumes the form $\frac{1}{\sqrt[]{2}}\begin{pmatrix}
1&i\\i&1
\end{pmatrix}$

In what follows, we develop the LLL formalism to describe dynamics and beam splitter features of single-particle propagation in the presence of a saddle potential. Following the single-particle description, where we show that the saddle Hamiltonian plays a role analogous to a squeezing operator in quantum optics, we turn to the two particle case, and identify the manner in which partitioning of these particles acutely depends on their fractional statistics. 

\subsubsection{Single Particle Dynamics}
Here, we discuss the time evolution of the coherent state postulated in previous sections in the presence of the saddle potential. The LLL coherent state structure enables us to develop the associated formalism in close analogy with quantum optics. 

Recall the Hamiltonian $H$ and angular momentum operator $L$ in the single-particle quantum Hall description in the absence of a potential landscape, as described in Sec II. In terms of the ladder operators, they are given by
\begin{align}
\mathcal{H}=\hbar\omega_0(B^\dagger B+1/2)&, L=\hbar(A^\dagger A - B^\dagger B)\\
iZ^*=A^\dagger-B&, -iZ=A-B^\dagger,
\end{align}
where $\omega$ is the cyclotron frequency and $Z$ is the complex position coordinate. The operators $A$ and $B$ commute, and hence, so do $Z$ and $Z^*$. 

The saddle potential featured in Fig.\ref{fig:saddlepot} generically has the form
\begin{align}
   H_S=UXY=i\frac{U\ell_b^2}{2}(Z^2-Z^{*2}), 
\end{align} 
where the strength of the saddle,  $Ul_B^2$, is assumed to be much smaller than the Landau level spacing, $\hbar \omega_0$. 

In what follows, we focus on the time-evolution of the LLL-projected coherent state defined in Eq. \ref{eq:coherent} under the influence of a saddle potential, namely $\ket{Z_i(t)}_c=e^{-iH_St}\ket{Z_i}_c$.On projecting to the LLL by applying the projection operator $P$, the operator $B$ becomes zero, as described by $BP=PB^\dagger=0$. Therefore, The Hamiltonian is effectively 
\begin{align}
    H_{S}=\frac{U\ell_b^2}{2i}({A^\dagger}^2-A^2)
\end{align}

In order to derive the transmission and reflection coefficients associated with the saddle potential, with regards to the physical process in Fig.\ref{fig:saddlepot}, we may consider a situation where a coherent state starts at an initial position close to the negative $X$-axis for small positive $Y$ far away from the scattering center. To obtain the coefficients, we may compare the time evolved state to a coherent state centered at a specific location in the complex plane. Towards this end, we define the following function:
\begin{align}
    f_Q(Z)=|{_c\bra{Z}\ket{Z_i(t)}}_c|^2\label{qfuncom}
    \end{align}
This distribution is called the Husimi Q-function. In the quantum optics literature, this distribution is extensively used to represent the wavefunction in phase space. Here, it represents the distribution of the coherent state in the lowest Landau level as time evolution takes place and provides a visual representation thereof.

In the particular situation of Fig.\ref{fig:saddlepot}, the transmission and reflection coefficients can be obtained by integrating the Husimi Q-function in the appropriate spatial regions. That is, we can define $F_R=\int_{Y>0}|{_c\bra{Z}\ket{Z_i(t)}}_c|^2 dZ dZ^*$ and $F_T=\int_{Y<0}|{_c\bra{Z}\ket{Z_i(t)}}_c|^2 dZ dZ^*$.

The fractions that propagate to the upper and lower half of the complex plane, respectively, are the parts of the wavefunction that are transmitted and reflected, i.e.
\begin{align}
   T=\frac{F_T}{F_T+F_R}\textnormal{, }R=\frac{F_R}{F_T+F_R} \label{tr}
\end{align}

In order to obtain these coefficients, we are required to evaluate the overlap between an arbitrary coherent state $\ket{Z}_c$ and the initial coherent state $\ket{Z_i(t=0)}_c$ time evolved under the influence of the saddle potential: 

\begin{align}
_c\bra{Z}e^{\frac{\xi}{2}(Z^2-Z^{*2})}\ket{Z_i}_c&={_c\bra{Z}}e^{\frac{\xi}{2}( A^2- A^{\dagger 2})}\ket{Z_i}_c, \textnormal{ }\xi=\frac{U\ell_B^2t}{\hbar}\label{eq:Singletime}
\end{align}

Key to our subsequent derivations, we observe that the time evolution operator takes the form of the squeezing operator, commonly used in quantum optics. In what follows, we thus present the formalism of squeezed coherent states, adapting techniques from quantum optics to the quantum Hall setting.  

\subsubsection{Squeezing Operators and Squeezed Coherent States}
Given a single mode system characterized by the ladder operators $A$ and $A^\dagger$, squeezing operators \cite{Lvovsky,Knight2004} are those that generate a Bogoliubov transformation in the space of these operators, preserving the commutation relations of the transformed operators. 
Squeezing operators take the generic form $S(\xi)=\exp{\frac{1}{2}(\xi^*A^2-\xi A^{\dagger2})}$, where we can parametrize the operator by $\xi=re^{i\theta}$.  Explicitly, their action on the creation-annihilation operators is
\begin{align}
c=S(\xi)AS(\xi)^\dagger&=A\cosh{r}-e^{i\theta}A^\dagger\sinh{r}\\
c^\dagger=S(\xi)A^\dagger S(\xi)^\dagger&=A^\dagger\cosh{r}-e^{-i\theta}A\sinh{r}\label{btrans}
\end{align}

Notably, the squeezing operators have a widely employed effect on coherent states. The resultant squeezed coherent states continue respecting the minimum Heisenberg uncertainty.  But the individual components of uncertainty in position $\Delta X$ and  momentum $\Delta P$ (for the quantum optics case) are different from those of regular coherent states. Physical generators of these squeezing operators are employed to reduce the uncertainty in one of the two conjugate variables.  In the quantum Hall situation, as discussed in previous sections, the conjugate variables are the coordinates $X$ and $Y$, and their uncertainty is dictated by the magnetic length.

Turning to squeezed coherent states, it is convenient to define the displacement operator parametrized by the variable $\alpha$, 
\begin{align}
    D(\alpha)=\exp{\alpha A^\dagger - \alpha^* A}.
\end{align}
A coherent state $\ket{\alpha}_c$ can be generated by the action of the displacement operator on the vacuum, namely $\ket{\alpha}_c =  D(\alpha)\ket{0}$. A squeezed coherent state is defined as $\ket{\alpha_\xi}_c=D(\alpha)S(\xi)\ket{0}$.  

With regards to the desired time evolution of the coherent state $\ket{Z_i}_c$ of Eq. \ref{eq:Singletime} subject to a saddle potential, we have
\begin{align}
\begin{split}
S(\xi)\ket{Z_i}_c&=S(\xi)D(Z_i)\ket{0}\\
&=D(\alpha)S(\xi)\ket{0}
\end{split}
\end{align}
where $\alpha=Z_i\cosh{r}-e^{-i\theta}{Z_i}^*\sinh{r}$. Just as the coherent state is an eigenstate of the annhilation operator, the squeezed coherent state is an eigenvector of the transformed operator $c$ as given in Eq. \ref{btrans}. Thus,  we can expand the squeezed states in the basis of the ladder operator modes (here, quantum Hall angular momentum states $\ket{k}$) to give \cite{Lvovsky}
\begin{align}
\ket{\alpha_\xi}_c=\sum\limits_{k=0}^\infty (k!\mu)^{-\frac{1}{2}}(\frac{\nu}{2\mu})^{\frac{k}{2}}\exp{-\frac{1}{2}(|\alpha|^2-\frac{\nu^*}{\mu}\alpha^2)}H_k(\frac{\alpha}{\sqrt[]{2\mu\nu}})\ket{k}\label{eq:sqstate}
\end{align}
where $\mu=\cosh{r}, \nu=e^{i\theta}\sinh{r}$ and $H_n(x)$ are the Hermite polynomials.

We note that the squeezed vacuum has the form
\begin{align}
\ket{0_\xi}=\frac{1}{\sqrt[]{\cosh{r}}}\sum\limits_{k=0}^\infty (-1)^k\frac{\sqrt[]{(2k)!}}{2^kk!}e^{ik\theta}(\tanh{r})^k\ket{2k}
\end{align}
Given the time-evolved state, we can use this squeeze operator formalism to evaluate the average position and uncertainties associated with this state. We find that given a coherent state at any initial position $Z_i$, we see that its average time-evolved coordinates follow an equipotential line along the saddle, given by 

\begin{align}
    (X_i e^{-\frac{U\ell_B^2t}{\hbar}},Y_i e^{\frac{U\ell_B^2t}{\hbar}}).\label{avgpos}
    \end{align}
    
Furthermore, the uncertainties in $X$ and $Y$ are given by
 
\begin{align}
\Delta X=\frac{l_B}{\sqrt{2}}e^{-\xi},\textnormal{ }\Delta Y=\frac{l_B}{\sqrt{2}}e^{\xi},\textnormal{ }\xi=\frac{U\ell_B^2t}{\hbar}
\end{align}

It is clear that the product of the uncertainties of the coherent state remains invariant on squeezing it, thus demonstrating the effect of the saddle potential as a generator of area preserving transformations. 

{\it Under the application of the saddle potential, a coherent state thus propagates along an equipotential line determined by its initial position and, in the process, becomes squeezed along a direction determined by the saddle parameters.}

For a coherent state originally located far from the origin along a negative value of the $x$ coordinate and slightly above the $X$-axis, as discussed above, the dynamics involves approaching the origin from the far left in Figure. \ref{fig:saddlepot} and being squeezed to reduce its width along $X$. Tunneling to the lower right quadrant occurs closest to the origin along its trajectory. A part of its wavefunction then transmits along the negative Y-axis and the remaining reflects along the positive Y-axis. 

In order to obtain the associated transmission and reflection coefficients defined in Eq. \ref{tr}, we need to evaluate $_c\bra{Z}S(\xi)D(Z_i)\ket{0}={_c\bra{Z}\ket{\alpha_\xi}}_c$. It follows from Eq.\ref{eq:sqstate} that the overlap has the form
\begin{align}
_c\bra{Z}\ket{\alpha_\xi}_c=\sum\limits_{n=0}^\infty \frac{{Z^*}^n}{n!}(\mu)^{-\frac{1}{2}}(\frac{\nu}{2\mu})^{\frac{n}{2}}\exp{-\frac{1}{2}(|\alpha|^2-\frac{\nu^*}{\mu}\alpha^2)}H_n(\frac{\alpha}{\sqrt[]{2\mu\nu}}). 
\end{align}
This overlap can be simplified using the identity 
\begin{align}
\sum\limits_{n=0}^\infty H_n(t)\frac{w^n}{n!}=e^{2tw-w^2}. 
\end{align}
Consequently, we obtain the closed form expression for the transmission and reflection amplitudes
\begin{align}
T(t)=\frac{1-\erf(\Gamma)}{2}\textnormal{, }
R(t)=\frac{1+\erf(\Gamma)}{2}\label{trvalues}
\end{align}
where $\Gamma=Y_i\sqrt{1+\tanh{\xi}}$. Here, recall that we have $\xi=\frac{U\ell_B^2t}{\hbar}$. Thus, for long times compared to $\frac{\hbar}{Ul_B^2}$, we have transmission and reflection coefficients that purely depend on the initial coordinate, $Y_i$. 

 This coherent state behavior is in marked contrast to transmission and reflection of energy eigenstates. For these eigenstates, the coefficients naturally depend on the specific equipotential line corresponding to the energy \cite{Halperin87,Stone}. For the saddle geometry, the closer the equipotential to the saddle point, the larger the energy-dependent tunneling (transmission) across the quadrant. Since equipotential lines respect the form $XY=C$, where $C$ is a constant, the coefficients depend on both the initial $X$ and $Y$ positions. 
 
The transmission behavior for coherent states, in its dependence purely on the initial $Y$ coordinate, is completely different from those of the energy eigenstates. A semi-classical picture provides intuition on this behavior of the transmission coefficient: the equations of motion are
\begin{align}
    i\hbar\frac{d \hat x}{dt}&=
[\hat X,H_{LLL}]=-iUl_B^2 \hat X \\
 i\hbar\frac{d \hat y}{dt}&=
[\hat x,H_{LLL}]=iUl_B^2 \hat Y
\end{align}
yielding the coherent state average position evolution given in Eq.\ref{avgpos}. Thus, the equations of motion for the two coordinates decouple. The transmission coefficient, which measures the net movement along the y-axis, only depends on the $Y$ coordinate. While this may seem surprising from the perspective of energy eigenstates, it is crucial to note that the components of the velocity of the coherent state are proportional to the coordinates, giving rise to very particular dynamics. As a result, for instance, for coherent states located along the same initial $Y_i$ position but different initial $X_i$ positions, the states further out from the origin travel faster. As a consequence, it can be shown that on average, each of these coherent states spends the same amount of time near the saddle point, giving rise to the same transmission coefficient.

Our discussion of single-particle saddle potential scattering and transmission is applicable to various quantum Hall bulk instances. For pinched geometries, our treatment provides a detailed description of tunneling across the pinched region through the bulk, which is usually modeled as a merely a phenomenological parameter when considering edge-state dynamics. Furthermore, the quantum Hall landscape is riddled by disorder and potential maxima and minima, as is crucial for understanding the integer quantum Hall properties. Tunneling between equipotential regions via saddle potential scattering is thus ubiquitous and is a key ingredient in descriptions such as the Chalker-Coddington network model. Our single-particle description is equally applicable for such quantum Hall bulk considerations. 
\begin{figure}
\centering
\includegraphics[width=0.4\textwidth]{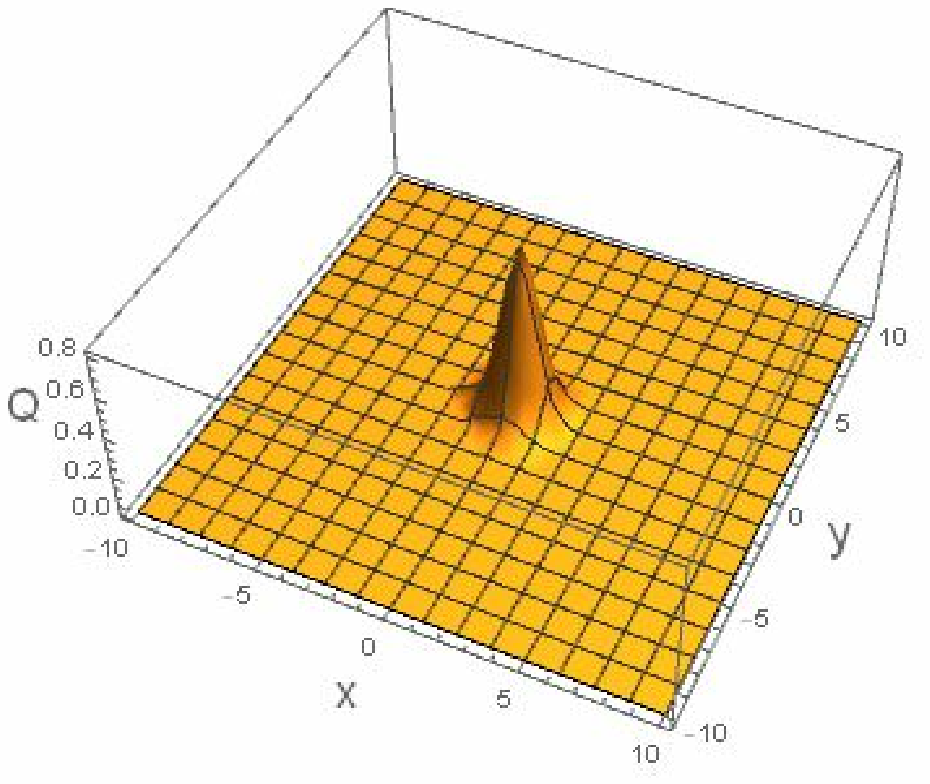}
\includegraphics[width=0.4\textwidth]{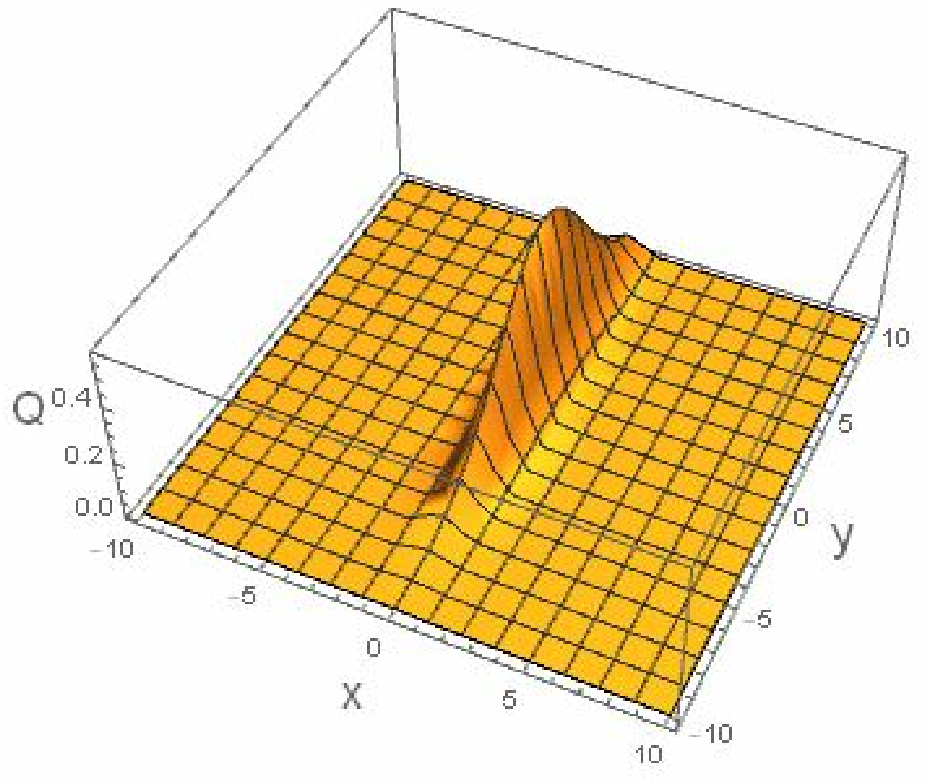}
\caption{\label{fig:SqSt} Depiction of a single-particle coherent state in the LLL in the Husimi Q-function representation of Eq.\ref{relQ}, which shows the magnitude of the overlap of the evolved coherent state with coherent states located at each point in the $x-y$ plane. The wavepacket is centered around an average initial position as shown in (a). When time-evolved under the action of a saddle potential, the coherent state gets squeezed by a time dependent squeeze parameter, as shown in (b). }
\end{figure}

\subsubsection{Two-Particle Coherent State Saddle Potential Dynamics}
In now analyzing two-particle evolution and effects of fractional statistics, the advantage with the saddle potential, as with the harmonic potential,  is its quadratic nature,  $H_S=\sum\limits_{\mu=1}^2 Ux_\mu y_\mu$.  We may thus once again separate out the potential into COM-relative coordinates. Projecting onto the LLL, the two-particle saddle Hamiltonian in this basis takes the form
\begin{align}
H_S=\frac{1}{2i}U\ell^2(A^{\dagger2}-A^2 + a^{\dagger2}-a^2),\label{saddleham}
\end{align}
where the COM operators $A$ and relative coordinate operators $a$ are as defined in earlier sections. 

Once more, two-particle coherent states associated with complex coordinates $z_1$ and $z_2$ can be decomposed into COM and relative coordinate spaces. Under time evolution due to the saddle potential in Eq.\ref{saddleham}, we have the extended version of the single-particle space given by
\begin{align}
\ket{Z(t)}_c&=e^{-\frac{Ut\ell_B^2}{2\hbar}(A^{\dagger2}-A^2)}\ket{Z_0}_c\\
\ket{\beta(t),\nu}_c&=e^{-\frac{Ut\ell_B^2}{2\hbar}(a^{\dagger2}-a^2)}\ket{\beta(0),\nu}_c
\end{align}
where $\ket{Z_0}_c$ is a coherent state in the COM coordinates centered at $(X,Y)$ and $\ket{\beta,\nu}_c$ is the relative coherent state centered at $(x,y)$.  

Hence, the COM behavior exactly parallels the single-particle situation. The time-evolved state $\ket{Z(t)}_c$ is a squeezed coherent state centered at $(X_0e^{-Ut\ell^2_B/\hbar},Y_0e^{Ut\ell^2_B/\hbar})$.  As with the single-particle discussion, we can now determine  explicit COM transmission and reflection probabilities for tunneling through the saddle potential in the LLL. These forms are identical to those of the single-particle case in Eq.\ref{trvalues}. 


For the relative coordinate, as a result of its explicit dependence on the statistical parameter, we find that a similar treatment proves to be much more subtle and resilient to analytic evaluation. In this sector too, we begin with the inner product between an arbitrary anyonic coherent state and the time evolved initial anyonic coherent state, 
\begin{align}
_c\bra{\beta,\nu}e^{-\frac{Ut\ell_B^2}{2\hbar}(a^{\dagger2}-a^2)}\ket{\beta_i,\nu}_c\label{relInPro}. 
\end{align}
We may once again evaluate this product using the Husimi Q-function, 
\begin{align}
|_c\bra{\beta,\nu}e^{\frac{\xi}{2}(a^2-{a^\dagger}^2)}\ket{\beta_i,\nu}_c|^2\label{relQ}
\end{align}
\begin{figure}
    \centering
    \includegraphics[width=0.45\textwidth]{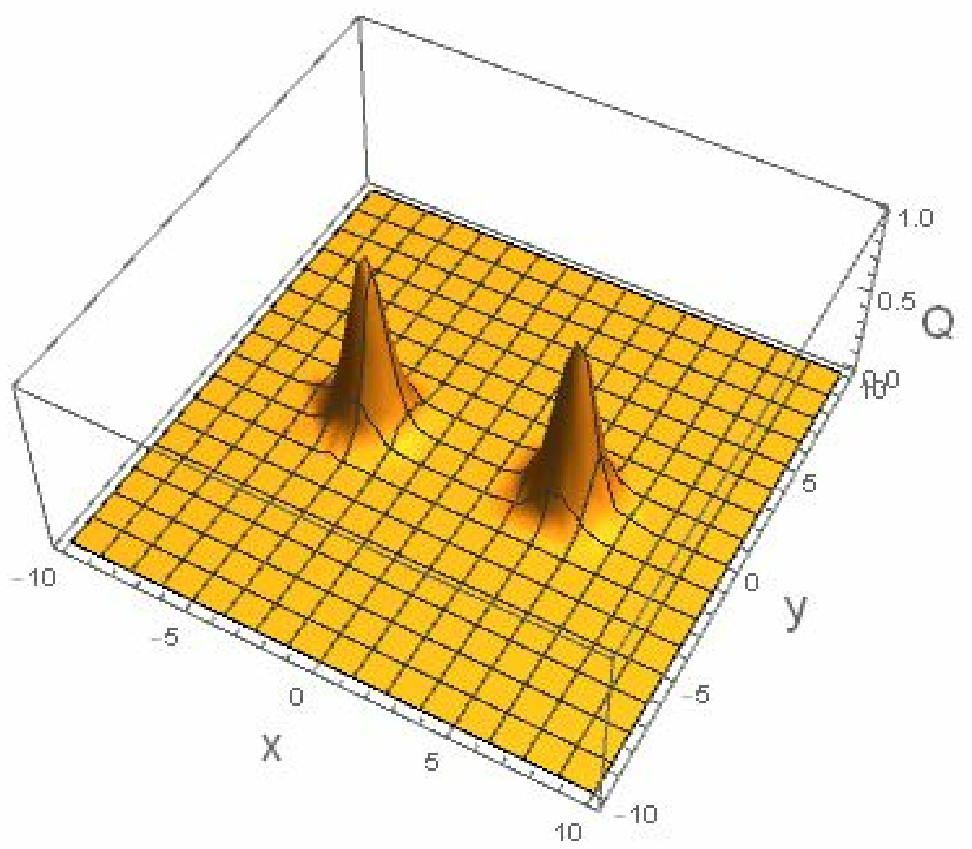}
    \includegraphics[width=0.45\textwidth]{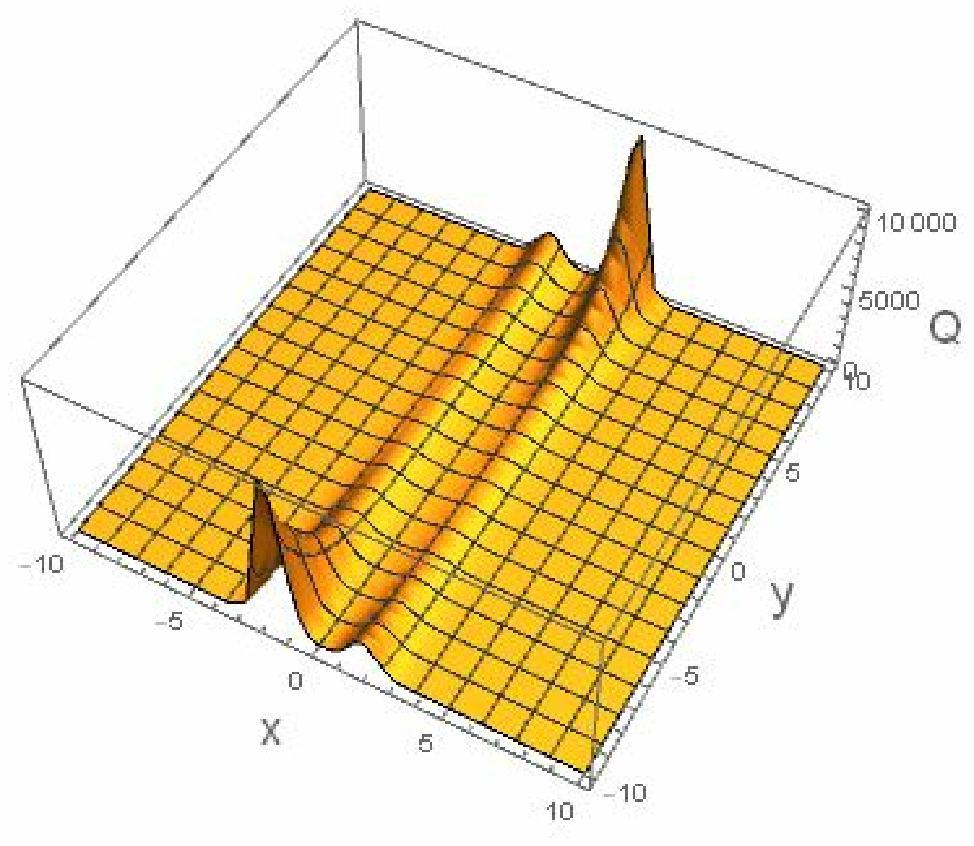}
    \caption{The relative coherent state for two particles in the LLL in the Husimi Q-function.  The relative coherent state is bimodal and symmetric about the origin, indicating its symmetry of indistinguishable particles, as shown for the initial state in (a).  When time-evolved under the action of a saddle potential, each coherent state gets squeezed by a time dependent squeeze parameter, as shown in (b), and splits into a transmitted and reflected wavepacket, while still maintaining the symmetry of the system.}
    \label{fig:relcohst}
\end{figure}
 To find this amplitude, we can resolve the time evolution operator using the decomposition theorems associated with the $\mathfrak{sp}(2,\mathbb{R})$ lie algebra described Sec.II (B). In particular, we express the time evolution operator as
 \begin{align}
  e^{(\frac{\xi}{2}(a^2-{a^\dagger}^2))}=e^{-\xi(B_+-B_-)}=e^{(-B_+\tanh{\xi})}e^{(-2\hat{J_1}\ln\cosh{\xi})}e^{(B_-\tanh{\xi})}
 \end{align} 
 
 Since the anyonic  coherent states are superpositions of generalized angular momentum states, and are eigenstates of the ladder operators $B_{\pm}$, the inner product in Eq. \ref{relInPro} is easily evaluated. The resulting infinite series can be written in closed form using the identity
 \begin{align}
     I_\mu(2x)=\sum\limits_{n=0}^\infty \frac{x^{2n+\mu}}{n!\Gamma(n+\mu+1)}
 \end{align}
 \begin{figure}
    \centering
     \includegraphics[width=\textwidth]{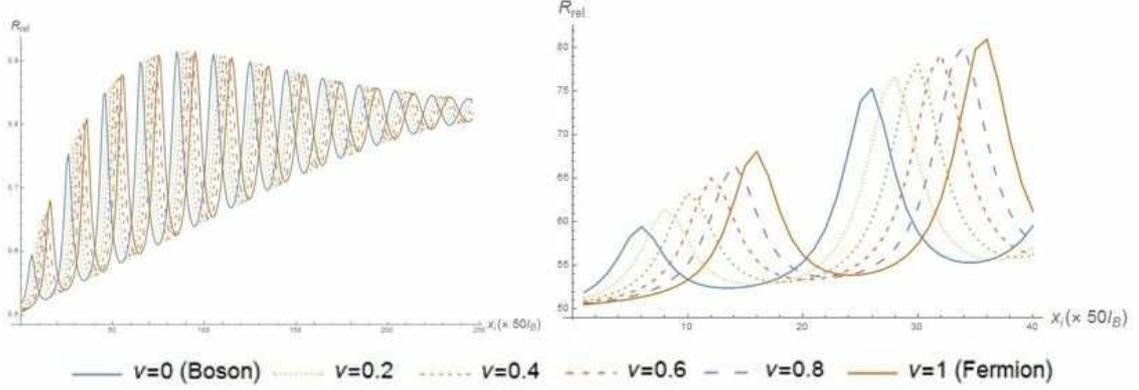}
    \caption{The reflection coefficient of the relative coherent state as a function of initial position along the x-axis,  plotted at a time $t=2.5\frac{Ul_B^2}{\hbar}$ and position $y_i=0.01 l_B$, well after the states have split into a transmitted and reflected wavepacket.  (b) Zoom of the graph shows distinctly distinct off-sets for different particles having different statistical parameters.}
    \label{TR}
\end{figure}
 where $I_\mu(x)$ is a modified Bessel function of the first kind. The Husimi Q-function thus obtained is plotted in the relative coordinate plane in Fig. \ref{fig:relcohst}. At all points of time, the relative coordinate coherent state is symmetric about the origin. This symmetry is but expected since we define this state as a representation of symmetries of quadratic functions in the LLL.  This specific constraint is a reflection of the more general property of indistinguishable particles that they not be told apart under exchange, as for instance with symmetrized and anti-symmetrized wavefunctions for bosons and fermions, respectively.

 Hence, upon time-evolution, each portion of the squeezed state  symmetrically splits up and moves along the y-axis, as seen in Fig.\ref{fig:relcohst}. As with the COM evolution, here too, upon impinging the origin, the wavefunction splits along transmitted and reflected direction, except that the symmetry due to indistinguishability is maintained. Therefore, to obtain the transmission (or equivalently reflection) coefficient, we can evaluate, for example, integrals over the quadrant $(x>0,y>0)$ and $(x>0,y<0)$ respectively. The Q-function itself is a ratio of Bessel functions and in deriving the reflection coefficient, its integral can only be evaluated numerically. Upon such integration, we obtain the coefficients as a function of $\nu$,  the initial position of the coherent state, and time. 
 
 The Fig \ref{TR} shows the reflection coefficient as function of initial relative x coordinate, $x_i$, for various statistical parameters. A small initial y coordinate $y_i$ is assumed. The plot depicts the coefficient at the particular time $t=\frac{2.5\hbar}{Ul_B^2}$ when the state has traveled well away from the origin and the split is well defined. Unlike in the single-particle case, the reflection coefficient does depend on the initial coherent states position $x_i$. As a general trend, we see that for small $x_i$, the transmission coefficient, and thus the probability of tunneling across the saddle is small. The coefficient then peaks at some optimal initial position along the $x$-axis, and finally tapers off to at some moderate value for large initial distance away from the original.  As the initial $y$ coordinate $y_i$ increases, the transmission gets smaller, and the peaks in the plot become less well-defined. The differences between the plots corresponding to different statistical parameters therefore become less significant before eventually merging into a common curve. 
 
  The second plot offers a closer view of the same graph. The plots show two key features: i) As expected, the transmission coefficient for coherent states corresponding to different statistical parameters $\nu$ is different though they start at the same initial position. ii) The plot shows regular peaks that are also $\nu$-dependent. This feature would be manifest in the analytic pole structure of the time evolution operator, or equivalently, the scattering matrix; further work is necessary to relate these features to the extensive analyses of the pole structure of the scattering matrix for the saddle potential\cite{}.

Returning to the behavior of the individual particles, we may now combine the results from the COM and relative coordinate analyses. Qualitatively, it is clear that two coherent states can be initialized at positions $z_1$ and $z_2$, and time-evolved along the contours of the saddle potential. Both particles impinge on the origin and split into transmitted and reflected parts in a manner that depends on the initial position, strength of the saddle potential, and the statistical parameter. The system thus acts as a two-particle beam-splitter. The overall transmission property can be characterized by the quantity
\begin{align}
    \int [dz_1][dz_2]|\bra{z_1\otimes z_2}S(t)\ket{z_{i1}\otimes z_{i2}}|^2&=\int [dZ][dz]|\bra{Z}S(t)\ket{Z_i}|^2|\bra{\beta,\nu}S(t)\ket{\beta_i,\nu}|^2\\
    &\sim T_{COM}T_{rel}(\nu).
\end{align}
Given initial positions, this quantity provides a measure of the manner in which the joint state is distribution in the complex plane upon time-evolution. It is worth emphasizing here again that the states are constructed by exploiting the quadratic symmetry of the $\mathfrak{sp}(2,\mathbb{R})$ Lie algebra appropriate to anyons. Hence, the relative state is always symmetric about the origin. Hence, making the states travel along the x axis will result in a 50-50 beam splitter irrespective of statistics.


To develop a further understanding of the nature of the time evolved coherent state, we now define a time-dependent dynamic version of the bunching parameter defined in Eq.\ref{bp1}:

\begin{align}
\chi(t)&=\frac{1}{4\ell^2}\Big[ {_c\bra{\beta(t),\nu}} \hat{r}^2\ket{\beta(t),\nu}_c- {_c\bra{z(t)_d}} \hat{r}^2\ket{z(t)_d}_c\Big]
\end{align}
The time evolution of the states under the saddle potential is again dictated by the relative coordinate Hamiltonian is $H=-i\frac{Ul_B^2}{2}({a^\dagger}^2-a^2)$.   We evaluate $e^{iHt} \hat{J_1}e^{-iHt}$ using the BCH formula and to obtain
\begin{align}
e^{iHt} \hat{J_1}e^{-iHt}=\hat{J_1}\cosh{2\xi}-\hat{J_2}\sinh{2\xi}, 
\end{align}
where the operator $\hat{J_2}$ corresponds to another boost operator, defined in Eq.\ref{gen}. 
Combining this identity with our treatment of the bunching parameter $\chi$ in Eq.\ref{bpe}, we obtain
\begin{align}
\chi(t)=\cosh{2\xi}(A_\phi-2|\beta|+\nu), 
\end{align}
where $\beta$ depends on the initial position of the coherent and the time dependence appears in $\xi=\frac{Ul_B^2}{\hbar}$. That is, on time evolution, the form of the bunching parameter does not change, it only gets multiplied by on overall time dependent function. This form indicates that the statistical nature of the particles is preserved after tunnelling through the saddle potential.

As discussed in Ref. \cite{Smitha10}, another useful indicator of the manner in which coherent state evolution is influenced by statistics  is the correlator between the $y$ coordinates of the two particles;  the saddle potential is such that the particles eventually travel along the y axis. This correlation has the form
\begin{align}
    \langle y_1 y_2 \rangle &= l_B^2e^{2Utl_B^2/\hbar}\Big(Y^2-\frac{y^2}{4}-\frac{\chi}{2}\Big)
\end{align}
The first two terms are the contribution of the coherent nature of the states in question, and would arise in distinguishable particle states as well. The third term increases or decreases correlations based on the sign of the bunching parameter, which in turn depends on the statistics of the particle in question. Thus,bosons show more correlation towards traversing in the same direction than fermions and anyons lie somewhere in between. Contrasting the behavior of distinguishable particles from indistinguishable cases in this saddle potential set-up thus enables direct extraction of bunching properties. 

To summarize the behavior of two-particle anyon LLL coherent states in the presence of quadratic potentials, we have modeled their dynamics when confined to harmonic traps. Particles encircle the minimum of the trap at a constant radius determined by their initial conditions. We explicitly show that in this situation anyonic signature naturally appears as a statistical phase due to either exchange or one particle encircling the other. In the case of the saddle potential, we have shown that coherent state dynamics corresponds to particles not only traveling along equipotential lines but also transforming to squeezed states. Furthermore, this saddle potential acts as beam splitter in that parts of the coherent state tunnel across the saddle point. For two particles, the dynamics is highly sensitive to the statistical parameter and beam splitter action directly reflects the bunching properties of these anyons. 


\section{Interferometry}

\begin{figure}
    \centering
    \includegraphics[width=0.7\textwidth]{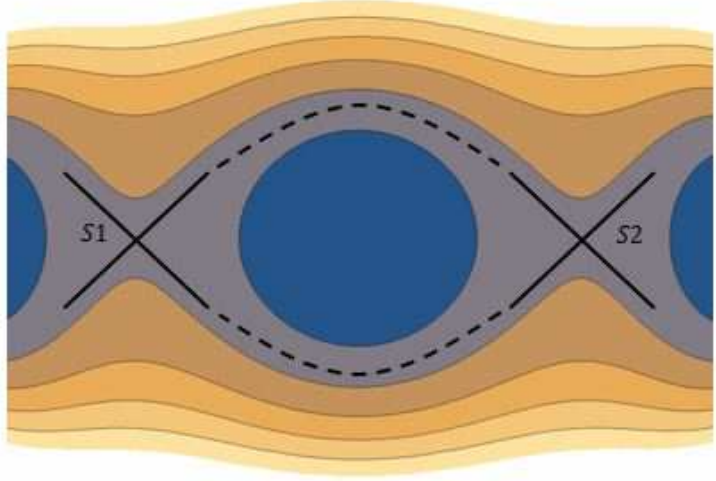}
   \caption{Contour plot denoting equipotential surfaces in a geometry inclusive of two saddle potentials flanking a harmonic trap. This setup mimics the bulk potential profile for quantum Hall geometries having two pinched regions and related interferometry properties.}
    \label{inter}
\end{figure}

 Anyon interferometry is a bustling area of study and extensive literature explores many aspects of it \cite{Cooper, Bonderson, Chamon}. Most theoretical studies approach anyon interferometry from an edge state tunnelling perspective and analyze fractional quantum Hall quasiparticles, as is natural in connecting with viable experiment \cite{Chamon, Eduardo, Stern, Chamon97, Willett2009, Fendley}. While these approaches, primarily employing Luttinger liquid edge state descriptions, are excellent in capturing salient features, ultimately a full-fledged understanding of the interferometry also needs considerations of quantum Hall bulk processes.  Here, we use the saddle and harmonic potential analyses above as building blocks for not only bridging LLL edge and bulk physics, but also designing a purely bulk-based interferometer. 
 


Consider an interference experiment carried out as shown in Fig \ref{inter}. In the edge state Hall case, this system corresponds to a Hall bar having two quantum point contacts created by pinched geometries. The quantum Hall bulk in a pinched region in fact experiences a saddle potential \cite{VanWees,Tekman, Buttiker90}. Typically, leads provide sources and drains at specific locations. A popular interferometry scheme involves injecting a quasiparticle at a given source $S1$ and measuring outputs at the drains. Tuning the magnetic field away from the center of the plateau enables controlling the number of quasiparticles in the bulk region between the two pinches. The injected quasiparticle can either traverse across the top edge undeflected, or backscatter at either point contact. Interference between two backscattered paths depends on both the Aharonov-Bohm flux enclosed by the two paths as well as any statistical phase acquired due to additional quasiparticles in this enclosed region. This setup also resembles standard Fabry-Perot interferometry\cite{FabPer}. 

The analyses of previous sections not only provide a bulk description of this interferometry, they offer the building blocks for various novel prospects for moving, manipulating, and interfering anyons in bulk settings. Specifically, the harmonic potential and the saddle potential combined open up a plethora of possibilities for initializing and dynamically evolving pairs of anyons. 

As an illustrative example, consider the situation for the interferometer described above. For bulk quasiparticles to mirror the edge-state quasiparticle in Fig.\ref{fig:saddlepot} (which in parts undergoes bulk tunneling), we may construct a bulk potential landscape as shown in Fig.\ref{inter}. While a full-fledged analysis of dynamics in this potential landscape would require numerical treatment, the salient features can be approximated by the saddle potential and the harmonic potential. Conforming to processes in Fig. \ref{inter}, the first process involves a single quasiparticle impinging the first pinch. This corresponds to our single-particle scattering against the saddle potential at $S1$. Next, the quasiparticle takes a curved trajectory to the second saddle, approximated by dynamics around a harmonic potential, which in principle could contain another anyon at its center. Subsequently, the quasiparticle scatters against the saddle potential at $S2$, a part of it scatters back, makes the return journey to complete its trajectory around the harmonic potential minimum, and finally impinges once again on saddle $S1$. Our analyses from previous sections for single-particle tunneling provides the relevant scattering processes at the saddle potentials. With regards to the quasiparticle encircling the middle portion, dynamics in the harmonic potential having an anyon at the center, also described in previous sections, provides two contributions. Time-evolution around an equipotential circle yields a phase factor attributed to the Aharonov-Bohm flux picked up in this region. The second factor arises due to the itinerant anyons encircling the static one. 

The example provided here is to demonstrate the first steps towards a formalism for describing bulk quasiparticle dynamics for a range of situations. The principle is to engineer appropriate potential landscapes that can be approximated by the quadratic landscapes for the two-particle situations studied here. Several such situations have been extensively explored in context of standard quantum Hall physics and edge-state geometries, without much attention given to the internal workings of bulk physics, which our study would fulfill as a complement. Examples include two-particle beam splitter geometries or Mach-Zehnder interferometry \cite{Stern,Bonderson,Feldman}. In principle, the formalism can even be extended to non-Abelian Ising anyons, as relevant, for instance, to $\nu=5/2$ quantum Hall states, by treating each fusion channel separately \cite{Stern, Willett2009,Yacoby}. Other situations entail a range from bulk phenomena in standard quantum Hall settings, such as Coulomb blockade physics due to trapped quasiparticles in local potential minima, to effective dynamics in photonic settings\cite{Ozawa,Simon}.

\section{Quantum Optics Analogies}
Our work heavily borrows formalism from the field of quantum optics and warrants a comparison between the parallels. The commonalities enabled us to derive several features of LLL particles. Differences in physical interpretation as well as new formalism that we built for treating unique anyon features, however, are also significant. 

The fundamental relationship that renders the parallel is the non-commutativity of position and momentum in regular quantum mechanics and of position operators in the lowest Landau level (LLL). The obvious differences are that the former is in phase space and non-commutativity is determined by Planck's constant while the latter is in real space and the non-commutativity is determined by magnetic length. In the former case, focusing on photons and oscillators, the infinite one-dimensional Hilbert space can be characterized by the number operator and corresponding energy levels. In the latter case, the infinite one-dimensional Hilbert space corresponds to degenerate LLL states, for instance, characterized by angular momentum in the symmetric gauge. Thus, in the former, energy level splitting gives rise to dynamics but not so in the latter. 

Non-commutativity in each case begs the identification of minumum uncertainty states. Corresponding coherent states respect mimumum uncertainty in a symmetric fashion for variables in both spaces. Coherent states can be thought of in multiple ways - i) as a minimum uncertainty state in the appropriate conjugate variable, ii) as an eigenstate to the annihilation operator, iii) and as the result of the displacement operator acting on the vacuum state. 
These different ways can be shown to be equivalent to eachother, and in a sense, these perspectives on the coherent state explain their ubiquitous applications. Coherent states, which form an overcomplete basis,  were originally studied by Schr\"{o}dinger and proposed as a solution to the correspondence principle - that is, in finding states that would show classical particle-like behaviour under the appropriate limits. These states were proposed in quantum optics as Glauber states, with seminal contributions to their understanding by Roy Glauber, Schwinger, ECG Sudarshan,  and several others \cite{Glauber,ECGS,Klauder:1960kt}. 

In the absence of an applied potential, in the photonic case, the coherent state executes circular motion in phase space while it remains fixed in real space in the LLL case. In both situations, further decreasing the uncertainty along one of the directions by increasing it along another can be rendered by the action of squeezing. In photonics, it is standard to perform a squeezing action through spontaneous parametric down conversion, which is a process of passing a higher energy photon through non linear optical materials to produce an effective squeezed state \cite{Lvovsky,Strekalov2019}.  These squeezed states are particularly useful in studying interferometry because they yield a smaller phase sensitivity of $\Delta\phi=\frac{1}{N}$ as opposed to a sensitivity of $\Delta\phi=\frac{1}{\sqrt{N}}$ for coherent states \cite{Yurke}. Optical coherent states of both the regular and squeezed varieties are constructed as superpositions of number states. The quantum optics formalism for the action of squeezing, particularly in the context of coherent states, offers a powerful, analytically tractable widely used description. Here, we have adapted this formalism for the parallel action of saddle potential dynamics in the LLL case and associated beam splitter physics. The appropriate states are superposition of angular momentum states. For the two-anyon situation, we have had to modify the formalism due to consideration of fractional angular momentum states and associated $\mathfrak{sp}(2,\mathbb{R})$ algebra. The action of a saddle potential on two-particle anyon coherent states, therefore, while having parallels with squeezing in quantum optics, exhibits significant deviations due to fractional statistics.

 Given the coherent state and squeezing formalism at hand, it is desirable to be able to visualize the states and their evolution in the appropriate space.  The Husimi Q-function elegantly offers one such visualization and is defined as $Q(z)=\frac{1}{\pi}\bra{z}_c\hat\rho\ket{z}_c$. Here, $\rho$ is a density matrix of the state being represented in a phase space described by coherent states $\{\ket{z}_c\}$. It is related to other phase space distributions, such as the oft-used Glauber-Sudarshan P-representation and the Wigner quasiprobability distribution. They are connected through the Weierstrass transform, which is an apodized transformation of the form -
\begin{align}
    Q(\alpha,\alpha^*)=\frac{1}{\pi}\int P(\beta,\beta^*)e^{-|\alpha-\beta|^2}d^2\beta=\frac{2}{\pi}\int W(\beta,\beta^*)e^{-2|\alpha-\beta|^2}d^2\beta
\end{align}
While our work exclusively focuses on the Husimi Q-function visualization, each of these forms has its own advantages. 

 The following table summarizes the analogies we draw between quantum optics and lowest Landau level quantum Hall systems
\begin{table}[H]
    \begin{tabular}{|l|l|}
    \hline
        \textbf{Quantum Optics} & \textbf{LLL Physics}\\\hline
        Spatial Dimensions: 1D & \multirow{2}{0.5\textwidth}{Spatial Dimensions: 2D, projected onto LLL to give an effective 1D} \\&\\\hline
        Defining Commutator: $[X,P]=i\hbar$ & Defining Commutator: $[R_x,R_y]=il_B^2$  \\\hline \multirow{2}{0.5\textwidth}{Ladder Operators $a,a^\dagger$ traverse between photon number states $\ket{n}$} & \multirow{2}{0.5\textwidth}{Ladder Operators $A,A^\dagger$ traverse between angular momentum states $\ket{k}$} \\&\\\hline
        Coherent State: $\ket{z}=e^{-|z|^2/2}\sum_{n=0}^\infty\frac{z^n}{\sqrt{n!}}\ket{n}$&  \multirow{2}{0.5\textwidth}{COM Coherent State: $\ket{Z}=e^{-|Z|^2/2}\sum_{k=0}^\infty\frac{Z^k}{\sqrt{k!}}\ket{k}$ Relative Coherent State: $\ket{\beta,\nu}_c=N_\beta\sum\limits_k\frac{\beta^k}{\sqrt[]{k!\Gamma(k+\nu+\frac{1}{2})}}\ket{k,\nu}$}\\&\\&\\&\\\hline
        Displacement Operator:$ D(\alpha)=\exp{\alpha a^\dagger - \alpha^* a}$& Displacement Operator:$ D(\alpha)=\exp{\alpha A^\dagger - \alpha^* A}$ \\&\\\hline Squeeze operator: $S(\xi)=\exp{\frac{1}{2}(\xi^*a^2-\xi a^{\dagger2})}$ & Squeeze operator: $S(\xi)=\exp{\frac{1}{2}(\xi^*A^2-\xi A^{\dagger2})}$ \\&\\ \hline
        \multirow{2}{0.5\textwidth}{The Husimi Q-function and the Wigner function are representations in phase space }& \multirow{2}{0.5\textwidth}{The Husimi Q-function and the Wigner function are representations in physical space}\\&\\ \hline
    \end{tabular}
    \caption{A tabulation of the analogies we draw with quantum optics}
    \label{tab:my_label}
\end{table}

Finally, with regards to the parallel, interferometry is fundamentally a tool for detecting wavefunction and geometric phases, and finds utility in a wide range of physics from optics to gravitational physics. It can also be readily adopted to perform electronic interferometry. As detailed in a previous section, edge state interferometry to study properties of fractional Hall states is well understood, and widely considered as a means of performing braiding operations on them. In this context, quantum point contacts behave as beam splitters, alongside engineered geometries that also enable beam splitting.  Historically, many of these concepts stem from the study of light. Quantum optics naturally offers both tools and language that can be readily adapted to elegantly describe dynamics of electronic and anyonic quantum states, with analogies working in both literal and metaphorical senses.

\section{The Inverted Harmonic Oscillator and Black Hole Parallels}
The problem of LLL (quasi-)particle scattering in a saddle potential not only has strong parallels with quantum optics, spatial non-commutativity gives rise to deep parallels in other contexts as well. In particular, the LLL projection reduces this two-dimensional setting to the problem of one-dimensional quantum scattering against an inverted harmonic oscillator potential. Surprisingly, the parallel also connects with phenomena related to black hole physics, in particular, Hawking-Unruh radiation and quasinormal modes. These parallels have been highlighted by us and co-workers in Ref. \cite{shortpaper}; here we offer a brief discussion of this study in the context of our current work.

The mapping to the inverted harmonic oscillator (IHO) is as follows. The Hamiltonian for the saddle potential in the LLL has the form $H_S=U(XY+YX)$, the $XY$ plane is now a non commutative plane.  The Hamiltonian can be rewritten in a canonically transformed basis as 
\begin{align}
    H=2U(P'^2-X'^2), \textnormal{ }P'=\frac{X+Y}{\sqrt{2}}\textnormal{ and }X'=\frac{X-Y}{\sqrt{2}}
\end{align}
where $[X',P']=-il_B^2$ respects the same commutation relation as $(X,Y)$. This structure of the lowest Landau level is comparable to a one dimensional phase space having an effective Planck's constant of $l_B^2$. The symplectic structure of phase space is thus also preserved in the LLL, and both spaces share invariance under transformations generated by the lie algebra $\mathfrak{sp}(2,\mathbb{R})$. Upon appropriate rescaling, we see that $P'$ can be identified with momentum in the one-dimensional situation and that the Hamiltonian above describes a particle of such momentum in the presence of an inverted oscillator potential. Thus, our treatment of coherent stated in a saddle potential in this current work effectively describes wave-packet scattering off an IHO.

The inverted harmonic oscillator is a simple, yet powerful and ubiquitous model in a wide range of phenomena from alpha-particle decay to metastability to Lyapunov behavior in chaos theory and Ads-CFT settings to inflation in the early Universe, and more. Its scattering properties have been well studied and play an important role in several of these phenomena. One such property is the thermal-like structure of the transmission amplitude of  energy eigenstates - 
\begin{align}
    T(\epsilon)=\frac{1}{1+e^{ \pi \epsilon/l_B^2U}}. 
\end{align}
In the context of the saddle potential, this amplitude can be obtained as Bogoliubov transformation\cite{Stone}.  It was pointed out in \cite{Stone} that this form resembles the thermal distribution obtained in the Hawking effect or the mathematically equivalent Unruh effect. 

In Ref.\cite{shortpaper}, we show that this resemblance to the Hawking-Unruh effect is more than a superficial coincidence, and is a reflection of the fundamental symmetries that govern these disparate physical systems. To elaborate, the Unruh effect can be viewed as the observation of a thermal distribution by a uniformly accelerating observer on measuring the vacuum state of an inertial or Minkowski observer. This effect arises because the spacetime of a uniformly accelerating observer is restricted to a section of the Minkowski spacetime called the Rindler wedge. This wedge is described by transformed space-time coordinates $(\tau,\zeta)$, where $t= e^{\zeta} \sinh(\tau), x=e^{\zeta} \cosh(\tau)$. This region of spacetime is left invariant by Lorentz boost transformations, generated by $\mathfrak{so}(2,1)$.  This Lie algebra is isomorphic to the $\mathfrak{sp}(2,\mathbb{R})$ algebra studied in this work. Intimately related to the saddle potential, the trajectory of the accelerating observer within this wedge is hyperbolic in nature. Just as evolution through a saddle potential gives rise to squeezing of states in the LLL, traversing through the hyperbolic trajectory makes the accelerating observer view the Minkowski vacuum as a squeezed vacuum. This squeezing action gives rise to the thermal distribution in the Unruh effect, and a thermal-like transmission probability in the case of scattering through a saddle. The IHO, being a generator of these algebras, appears in both contexts, as a squeezing or shearing operation in the LLL on the one hand and as a boost in Minkowski spacetime on the other hand. In this sense, it also generates time translation in the frame of the accelerating observer (since the accelerating observer is constantly being boosted) and simply serves as the Hamiltonian for dynamics within the Rindler wedge.  In this way, we now have a parallel to Lorentz kinematics playing out in a non-relativistic quantum arena.

A powerful prediction from the IHO and black hole perspectives is the existence of so-called quasinormal modes. Originally predicted in the black hole context by C. V. Vishveshwara \cite{Vish}, these resonant modes occur due to scattering by a wavepacket of finite width in energy against a potential maximum and are manifestations of purely outgoing boundary conditions\cite{Bohm, Perelomov,CHRUSCINSKI}. They decay in time, have a finite amplitude at the system's boundaries, and are ideal for modeling processes involving net current leaving a system, such as with alpha- particle decay. In black holes, these quasinormal modes are associated with signature gravitational wave black hole signals, which in the spherically symmetric Schwarzschild case purely depend on the mass of the black hole. Recently, these modes have been invoked in the context of black hole merger ringdown signals detected by LIGO \cite{LIGO1}. The mapping to the LLL saddle situation and related quantum Hall settings  provides an arena to probe these decaying modes. In \cite{shortpaper}, we propose a physical experiment to realize quasinormal modes by means of Gaussian scattering across a saddle potential. These modes would appear as time decaying oscillations in the reflected and transmitted portions of the wavepacket. The existence of these modes is encoded in the analytic properties of the scattering matrix associated with the IHO, and equivalently, in the time evolution operator. In future work, based on results of the current work, we propose to explore the connection between  the time evolution operator as a squeezing operator, coherent state dynamics, and quasinormal mode physics. Most importantly, the parallel not only provides fertile ground for exploring gravitational phenomena in LLL settings and vice-versa, as with our proposed quasinormal mode probe through quantum Hall point contact time-resolved measurements, lesson from one field can provide completely novel predictions in the other.

\section{Summary and Future Directions}

In summary, we have explored correlations and statistical properties of lowest Landau level particles endowed with fractional statistics and their dynamics under the influence of shallow potential landscapes. Anyon dynamics is especially important in bringing out exotic topological properties and manipulating fractional particles to obtain signatures and physical applications of these properties. Physically, such particles are most successfully realized in fractional quantum Hall systems. Though the anyonic excitations in quantum Hall systems are a result of many-body interactions, they can effectively be modelled as localized states in the system. We have used this feature to model the anyons as coherent states. Coherent states are particularly effective in mimicking particle behaviour because they are minimum uncertainty states, are localized in the LLL, and their dynamics can be approximated by semi classical means. By picking the appropriate Lie algebra that represents the symmetries inherent to anyons, we construct a two body anyonic coherent state. 

In this comprehensive study, we have drawn attention to the behaviour of quantum Hall bulk Abelian anyons in a harmonic trap and in the presence of a saddle potential.  The fractional statistics associated with the anyons have been incorporated by means of a statistical phase $\pi\nu$ that is picked up upon exchange of the anyon pair.  We have shown that on time evolution in the presence of a harmonic potential, the coherent states move in circular trajectories. On completing a full circle, the anyonic states acquire an overall phase of $2\pi\nu$. We have demonstrated that time evolution through a saddle potential is effectively a problem of wavepacket scattering against an inverted harmonic potential. Borrowing from quantum optics formalism, we have found that the action of this potential is that of a squeezing operator having a time dependent parameter. Hence, coherent states tunnel through the saddle point while getting squeezed on time evolution. We have estimated the reflection and transmission coefficients in terms of the initial position of the two particles. Thus, the saddle potential acts like a beam splitter that distinguishes between particles of different statistics through differences in the transmission and reflection coefficients. As an example of the utility of such dynamics as building blocks in more complex geometries, we have demonstrated how a configuration of hamronic and saddle potentials can create an anyonic interferometer. In future work, we aim to extend this formalism to non-Abelian anyons and employ such geometries to observe braiding and other non-Abelian effects.

Our findings are relevant to a range of physical settings concerning quantum Hall physics and anyons. At a fundamental level, we have shown bunching and exchange properties that are common to Abelian anyons in general. Lowest Landau level coherent state dynamics in harmonic and saddle potentials are generic enough for any quantum Hall bulk situation. As specific realizations, in solid state systems, be it semiconductor-based materials extensively studied over decades or new topological and graphene-based materials \cite{Falko,Novoselov1379,Jacak2017},  while much of the focus is on edge-states, bulk physics too is of significant interest and is becoming more accessible with the development of novel experimental probes. In the context of our results, for instance, previous work identifying fractional quasiparticles trapped in local potential minima through Coulomb blockade measurements \cite{Martin980,Patton_2014}, can potentially be extended to include patterned minima, saddle potential based beam-splitters and bulk interferometers, as well as detectors, such as SETs, for correlated measurements. Lowest Landau level physics has also enjoyed attention in several new viable settings, such as cold atomic systems experiencing rotation or synthetic gauge fields and synthetic photonics-based topological materials \cite{Simon,Jacob_2008,Sorensen,Ozawa,Hafezi2013}; these settings enable highly controlled dynamic application of potentials and manipulation of wavepackets and would thus be excellently matched to investigate the physics studied here. 
 
 While the dynamics presented here is restricted to the two-anyon case, it offers a glimpse into the drastically different non-equilibrium behavior of many-particle systems, when compared to fermions and bosons. Extensions beyond two-particle analyses could benefit from formalisms employing flux attachment, such as in Chern-Simons theories, or statistical treatments, such as with anyon gases \cite{textbook,leshouches,MikeStone,Prange,Jain}. However, even restricting to two particles, headway can be made in non-equilibrium situations, such as quench dynamics coming from dynamically tuning a parameter in the governing Hamiltonian. Past work has emphasized distinct differences in quench behavior stemming from topological order \cite{Sondhi1,Hegde_2015,Sondhi2,Kells}, associated ground state degeneracies and anyonic excitations. Considerations of the saddle potential dynamics become relevant to such situations when, for instance, the quench involves tuning the magnetic field and associated filling fraction in quantum Hall systems, thus nucleating new quasiparticles or altering the effective potential landscape. Other quench scenarios have analyzed dynamically altering the tunneling amplitude in point contact settings and studying the growth of entanglement entropy \cite{Fradkin_2009,Hsu}.  The work presented here paves the path for preliminary work on such non-equilibrium quench dynamics in quantum Hall systems. 

Lowest Landau level physics and saddle potential dynamics, in their elegance and simplicity, find parallels in diverse branches of physics. At heart, the commonality stems from the non-commutative nature of the lowest Landau level and the ubiquitous nature of the inverted harmonic potentials. While we have extensively discussed the parallels with quantum optics, we have barely touched those with quantum condensed phases of matter. In particular, the commonality lies in hyperbolic transformations and Bogoliubov excitations, which naturally arise in the condensate context. They also are integral to the Unruh effect, naturally linking with spacetime geometry and Hawking radiation. As a more palpable non-equilibrium black hole phenomenon, wave packet dynamics in the lowest Landau level can simulate quasinormal modes characteristic of these enigmatic objects excited in cataclysmic events, such as black hole mergers. In conclusion, rich parallels to the work presented here extend across a range of fields from quantum optics to gravity and are ripe for further investigations.

\section{Acknowledgements}
We are grateful to Barry Bradlyn, Suraj Hegde, and Michael Stone for their illuminating discussions. This work is supported by the National Science Foundation under Grant No. 1745304 EAGER:BRAIDING.

\bibliography{ref}
\end{document}